\documentclass[prc,reprint,superscriptaddress,floatfix,onecolumn,notitlepage]{revtex4-1}

\usepackage{geometry}                
\usepackage{graphicx}
\usepackage{amsfonts}
\usepackage{amssymb}
\usepackage{amsmath}
\usepackage{epstopdf}
\usepackage{colortbl} 
\usepackage{multirow} 
\usepackage{lineno} 
\usepackage{setspace} 
\usepackage{fullpage} 
\usepackage{fancyhdr}
\usepackage{color}
\usepackage{booktabs,longtable,tabu} 

\definecolor{green}{rgb}{0,0.3,0}

\usepackage[toc,page,title]{appendix}

\newcommand{\un}[1]{\ensuremath{\ \mathrm{#1}}} 

\makeatletter
\def\l@subsubsection#1#2{} 
\makeatother

\usepackage[pdftex]{hyperref} 

\begin{document}

\title{Scientific Report: The Systematics and Operational Studies (SOS) Apparatus as a testbed for nEDM@SNS experiment}


\maketitle

\begingroup
\centering

 written by V. Cianciolo$^1$, R. Golub$^{2,3}$, B. W. Filippone$^3$, P. R. Huffman$^{2,3}$, K. Leung$^{2,3}$, E. Korobkina$^{2,3}$, C. Swank$^{2,3,4}$ on behalf of the nEDM@SNS collaboration

{\textit{$^1$Oak Ridge National Lab, TN, USA}}

\textit{$^2$North Carolina State University, Raleigh, NC 27695, USA}

\textit{$^3$Triangle Universities Nuclear Laboratory, Durham, NC 27708, USA}

\textit{$^4$California Institute of Technology, Pasadena, CA 91125, USA}

\endgroup

\noappendicestocpagenum
\tableofcontents
\addtocontents{toc}{\protect\setcounter{tocdepth}{1}}
\section{Introduction}

The nEDM experiment at the SNS (nEDM@SNS)~\cite{ahmed2019new} is the first measurement of neutron EDM to directly measure the precession frequency of the neutron spin due to magnetic and electric fields. Previous measurements have inferred the precession frequency by measuring the residual polarization of neutrons after a long period of free precession. This difference provides independent information on potential unknown systematic uncertainties compared to previous and ongoing measurements. It is precisely because nEDM@SNS is using a number of novel techniques that it is essential to perform detailed studies of these techniques in order to optimize the statistical sensitivity and minimize the systematic uncertainty. Although these studies can and in some cases will be done with the full-scale apparatus, the significant cost of operating the nEDM@SNS (including the significant down-time required to make small changes to the apparatus) drives us to consider alternative, cost-effective ways to perform some of these studies.
 
The nEDM@SNS apparatus is designed to maximize statistical sensitivity and minimize systematic uncertainty in extracting a neutron EDM signal. This mandates long neutron storage times as well as long neutron and $^3$He polarization times. These in turn require highly uniform fields and high $^3$He polarization at a low concentration. The Systematics and Operational Studies Apparatus (SOSA) does not need this level of statistical sensitivity and systematic error reduction. Therefore, it can operate with less field uniformity and lower $^3$He polarization. This allows the SOSA to be a much more nimble device that can have a turn around time of a few weeks rather than months, which in turn will be much less expensive to operate.
 

We identify three major studies to be performed with the SOS apparatus:
\begin{itemize}
\item Detailed studies of neutron and $^3$He correlation functions to characterize key systematic uncertainties
\item Optimization of simultaneous spin manipulation of polarized neutrons and $^3$He
\item Identification of high-performance measurement cells
\end{itemize}

 

\section{Overview of the Systematics and Operational Studies (SOS) Apparatus}

A Systematics and Operational Studies (SOS) apparatus has been designed to perform measurements that require the presence of polarized ultracold neutrons (UCNs) and/or polarized $^3$He dissolved in superfluid helium near 400~mK. There will be magnetic fields for spin manipulation and two detection techniques: SQUIDs for $^3$He magnetization measurements, and scintillation light detection to monitor UCN density and the angle between the $^3$He and UCN spins. There is no need for electric fields.

The total volume of LHe in the SOS apparatus is approximately 5 liters and the turnaround time of the system (cooldown \& warm up) is $\sim$ 2 weeks ($\sim$ 4-5 times shorter than nEDM@SNS). Polarized $^3$He will be injected and removed each cycle with a turnaround time of $\sim$ 2-3 hours ($\sim$ 10 times longer than nEDM@SNS). Thus, while the main nEDM@SNS apparatus is more effective in collecting high-sensitivity data, the SOS apparatus is preferable for commissioning phase measurements that require frequent warm-up and opening for adjustments and, likely, many more cool-downs.

\begin{figure}[htbp]
\begin{center}
\includegraphics[width=10cm]{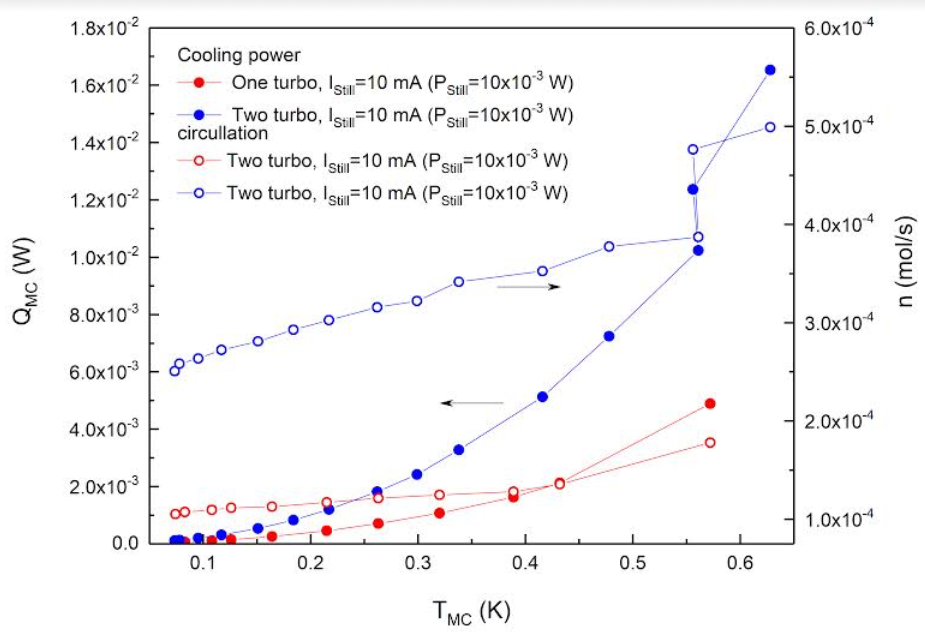}
\caption{Temperature dependence of Cooling power and He-3 circulation flows of the SOS Dilution Refrigerator. Solid circles - one pump, open circles - two pumps}
\label{fig:CoolingPower}
\end{center}
\end{figure}

A simplified schematic of the SOS apparatus is shown in Fig.~\ref{fig:simplifiedPULSTARschematic} with a more detailed description and figures found in Appendix~\ref{appen:overviewApparatus}. The cooling system is based on the use of an existing Dilution Refrigerator (DR) that has a cooling power of $\sim 5\un{mW}$ at 400\un{mK} ( see ~\ref{fig:CoolingPower}). Originally, DR had only one pump, but we have added another one to increase the cooling power. The DR mixing chamber will be connected to the $^4$He-buffer/$^3$He-evaporation cell, which in turn is linked to the measurement cell through 0.5'' tubing filled with superfluid $^4$He serving as a cooling link. During cooldown, the isotopically pure $^4$He will be condensed into the buffer cell.

\begin{figure}[htbp]
\begin{center}
\includegraphics[width = 16cm]{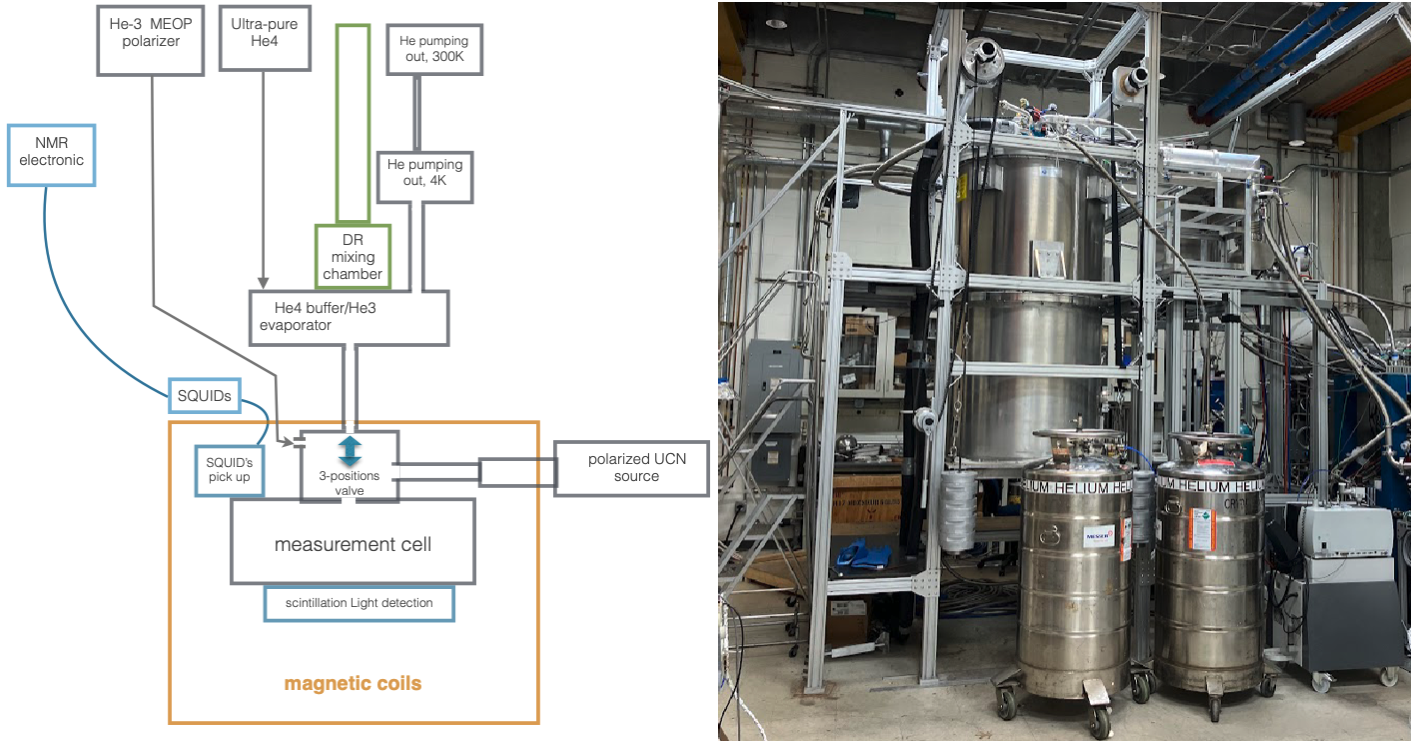}
\caption{Left - Simplified layout of the Systematics and Operational Studies (SOS) apparatus at PULSTAR. Note that in Phase I (studies with only polarized He-3) there will be no UCN guide and light collection system. Right - photo of the Phase I of the SOS apparatus at TUNL, Duke University}
\label{fig:simplifiedPULSTARschematic}
\end{center}
\end{figure}

Isotopically clean $^3$He will be polarized using the Metastability Exchange Optical Pumping (MEOP) technique for $\sim 10\un{min}$ at a pressure of a few mbar to achieve a $^3$He polarization up to 70\%. The required amount of $^3$He atoms will be diluted to achieve the desired $^3$He density in the cell, mixed with $^4$He to $\sim 500\un{mbar}$, and then injected to the measurement cell via cryopumping onto the liquid surface inside the apparatus. The $^3$He will then transport into the measurement cell via diffusion.
 
Removal of depolarized $^3$He from the measurement cell at the end of cycle will be done with the heat flush technique, which uses phonon flow (produced by heating the cell) to force $^3$He into the evaporator cell. The $^3$He vapor will be then pumped out by a charcoal adsorption pump at 4\un{K}. The pump can be regenerated when saturated by heating and then pumping out with an external room temperature pumping system.

Neutrons will be delivered to the cell through the 3-position ``vestibule'' valve from an external UCN source. The current plan is to use the PULSTAR UCN source, which is now at the final stages of cryogenic commissioning. The back-up plan is to use the LANL UCN source. 

The measurement cell will be made from PMMA as is the case for the nEDM@SNS experiment. The cell can be made identical to the nEDM cell or smaller for additional studies. The 3-position valve housing will have a flange identical to the nEDM cell to allow testing of the nEDM cells before assembling them into the nEDM apparatus. The 3-position valve will allow: (a) filling the measurement cell with polarized $^3$He and UCN, (b) storing UCN and $^3$He, and (c) connection of the measurement cell to the evaporator cell for removal of depolarized $^3$He.

The neutron absorption cross section of $^3$He is strongly dependent on the angle between the spins of the neutron and $^3$He: it is nearly zero when the spins are parallel and is at a maximum of 11\un{kilobarns} at thermal neutron energies when the spins are anti-parallel. The angle between the neutron and $^3$He spins can be monitored by the rate of n-$^3$He capture resulting in scintillation light produced in the superfluid helium. To detect the light, we will use and test a prototype light collection system developed for the nEDM@SNS experiment. It will utilize wavelength shifting (WLS) fibers coupled to Silicon Photomultipliers (SiPMs) located at 77~K. The $^3$He magnetization will be monitored by a SQUID magnetometry system using the latest design (e.g. the currently envisioned``4 pack'' configuration) and its associated electronics for the nEDM experiment.

The magnetic fields required for spin related experiments will be produced by a magnetic coil package, designed in a similar way as the magnetic package of the nEDM@SNS experiment, but with less stringent field gradient requirements to reduce the size of the SOS apparatus coil package. This approach will allow us to study different procedures for spin manipulation and check if any hardware adjustments are required to the full-scale nEDM magnet package in advance to assembling of the main nEDM apparatus.

\section{Scientific Program of Measurements with SOS apparatus \label{sec:keymeasurements}}

 

The timeline of our studies are divided into two phases. In Phase I, the polarized $^3$He injection and removal system, SQUID system, and magnetic field coils will be installed. This will allow measurements involving polarized $^3$He (e.g. $^3$He friendliness of cells and $^3$He correlation studies). In Phase II, the apparatus will be coupled to a polarized ultracold neutron source and the light collection system installed. This will allow the rest of the measurement program (e.g. neutron friendliness of cells and simultaneous spin manipulation of the two species) to be performed.

\subsection{Control of systematic effects via measurement of correlation functions \label{sec:correlationFunc}}

The leading systematic effect in the nEDM@SNS experiment is the false EDM produced by the interaction between the $\vec{E} \times \vec{v}/c$ motion-induced magnetic field and ever-present magnetic field gradients. The combination of these two fields produces time-dependent magnetic fields in the plane perpendicular to $B_0$, and thus cause Bloch-Siegart precession frequency shifts that turn out to be linear in $E$ (and flip sign with the direction of E). This is often called the ``geometric phase induced'' frequency shift, but will be called the ``linear-in-E'' shift here, and denoted by $\delta \omega$. The size of this shift depends on the geometry of the trap and the magnetic field gradients present. Furthermore, it also depends on a particle's gyromagnetic ratio, velocity and trajectory, so that the size of $\delta \omega$ is different between the neutron and $^3$He. Therefore, this shift is generally not corrected for by the $^3$He comagnetometer and will produce a false neutron EDM signal.

The philosophy of the measurement program described here is to measure the correlation functions of the $^3$He and ultracold neutron motion via spin relaxation times (or frequency shifts) induced by the presence of known applied magnetic field gradients. The electric field is not required since it does not affect the neutron or $^3$He motion. Once the correlation functions are determined, they can be used to predict and control the systematic error caused by the linear-in-E frequency shift and its associated false EDM effect. Indeed, for $^3$He the size of $\delta\omega$ can be made minimal by an appropriately tuned superfluid helium temperature because the mean-free-path of $^3$He diffusion, dominated by $^3$He scattering off phonons in the superfluid, is strongly temperature dependent ($\propto T^{7.5}$). The ultracold neutron motion is unaffected by temperature since the phonon scattering time constant at $0.5\un{K}$ is $\gtrsim 10,000\un{s}$, and if they were to scatter, they would be lost since the typical energy gain is much larger than the storable UCN energy.

The variation of the B-field over the measurement cells can be characterized by the spectrum of the correlation function \cite{redfield,Lam05,pignol}:
\begin{align}
S_{B_iB_i}\left( \omega_0 \right)  =\int_{0}^{\infty }\left\langle B_{i}(t)B_{i}(t+\tau )\right\rangle e^{-i\omega_0 \tau }d\tau \;,
\end{align}
where $\omega_0=\gamma B_0$, $B_i$ is the field along $\hat{i}$ experienced by the particle, and $\left<..\right>$ is the ensemble average. 

With the holding field $B_0$ along the $z$-direction, the $x$-direction as the long, 40~cm axis of the cell, and the $y$-direction as the 10.2~cm dimension of the cell, the linear-in-E frequency shift $\delta \omega$ is written as:
\begin{align}
\delta\omega =\frac{\gamma ^{2} E}{c^2}\left[\omega_0
\mathrm{Im}\left(S_{B_xB_x}(\omega_0 )\right)+\omega_0\mathrm{Im}\left(S_{B_yB_y}(\omega_0 )\right)+<xB_x>+<yB_y>\right] \;.
\label{eq:dw}
\end{align}
A sensitive measurement of the correlation function allows for characterization and potentially subtraction of the linear-in-E shift. Therefore, a quantitative understanding of the behavior of the correlation functions for both $^3$He and neutrons near the expected operating parameters is critical.  

A quantitative characterization of the imaginary part of the correlation function allows for correction of $\delta\omega$. In fact the correlation functions can be accessed without an E-field by studying spin relaxation. For example in a constant applied magnetic gradient, $G_i={\partial B_i}/{\partial i}$, the longitudinal relaxation is given by:
\begin{align}
\frac{1}{T_{\rm 1,gradient}} =\frac{\gamma ^{2}}{2}
\left(G_{x}^{2}\mathrm{Re}\left[S_{xx}(\omega_0 )\right]+G_{y}^{2}\mathrm{Re}\left[S_{yy}(\omega_0 )\right]\right), \label{eq:T1}
\end{align}
where, for example, $S_{xx}\left( \omega_0 \right)  =\int_{0}^{\infty}\left\langle x(t)x(t+\tau )\right\rangle e^{-i\omega_0 \tau }d\tau$. 

The transverse relaxation for an applied gradient oscillating with frequency $\omega_{\rm rf}$ is given by:
\begin{align}
\frac{1}{T_{\rm 2,rf\,gradient}} =\frac{\gamma ^{2}G_{zx}^{2}}{2}\mathrm{Re}\left[S_{xx}(\omega _{\rm rf})\right] \;,\label{eq:T2}
\end{align}
where $G_{zx} = {\partial B_z}/{\partial x}$.

While Eqs.~\ref{eq:T1} and \ref{eq:T2} are in terms of the real part, Eq. \ref{eq:dw} depends on the imaginary part. In principle, the real and imaginary parts (cosine and sine transform) are related by the integral equations:
\begin{align}
\mathrm{Re}\left[ S_{xx}\left( \omega \right) \right] &=\int_{0}^{\infty }R_{xx}(\tau)\cos (\omega \tau)d\tau \label{eq:ctrans} \textrm{ , and }\\
\mathrm{Im}\left[ S_{xx}\left( \omega \right) \right] &=\int_{0}^{\infty }R_{xx}(\tau)\sin \left( \omega \tau \right) d\tau =\frac{1}{2\pi }\int_{-\infty }^{\infty }\frac{\mathrm{Re}\left[ S_{xx}(\omega ^{\prime })\right] }{\omega -\omega ^{\prime }}d\omega ^{\prime } \;,
\label{eq:strans}
\end{align}
where
\begin{align}
R_{xx}(\tau)&=\left\langle x(t)x(t+\tau )\right\rangle 
\end{align}
is the position auto-correlation function. Due to experimental constraints, it is not feasible to measure the correlation function across a sufficient range of $\omega$ to perform the integral in Eq.~\ref{eq:strans}. However, the correlation function is determined from a conditional density $p(x,x_{0},\tau )$, given by,
\begin{align}
S_{xx}\left( \omega \right) =\int_{-\frac{L}{2}}^{\frac{L}{2}}\int_{-\frac{L}{2}}^{\frac{L}{2}}xx_{0}p(x,x_{0},\tau )e^{-i\omega \tau }dxdx_{0}d\tau 
\end{align}
so that we can fit the results to a model dependent conditional density that can be derived from both theory \cite{Swank2016} and simulation \cite{riccardoThesis}. The resulting conditional density will accurately predict the linear-in-E phase shift.

The typical gradients for measuring $T_{\rm 1,^3He,gradient}$ to determine the $^3$He correlation function are $\sim 200 \un{\mu G\,cm^{-1}}$, producing $T_{\rm 1,^3He,gradient} \sim 100{\rm\, s}$ (see Fig.~\ref{fig:T2relaxation} of Appendix~\ref{appen:corrFunc}). Due to the large polarized $^3$He concentration $x_{\rm pol\,3} \sim 10^{-7}$ that can be used without changing the motion of the $^3$He (i.e. the $^3$He mean-free-path remains dominated by scattering off phonons in the superfluid helium), reaching the required sensitivity can be easily achieved. The expected precision on the correlation function extracted from this technique in the worst-case is $\lesssim 10\%$, but will likely be much better after gaining experience on implementing the techniques. Measurements of $T_{\rm 1,gradient}$ to extract the correlation functions in small $\sim1\,{\rm in^3}$ cells and at different $^3$He motion regimes (due to the high $x_{\rm pol\,3}$ required for sufficient signal) have already been demonstrated in separate measurements~\cite{swankThesis}.

\begin{figure}[htbp]
\begin{center}
\includegraphics[width=.7\textwidth]{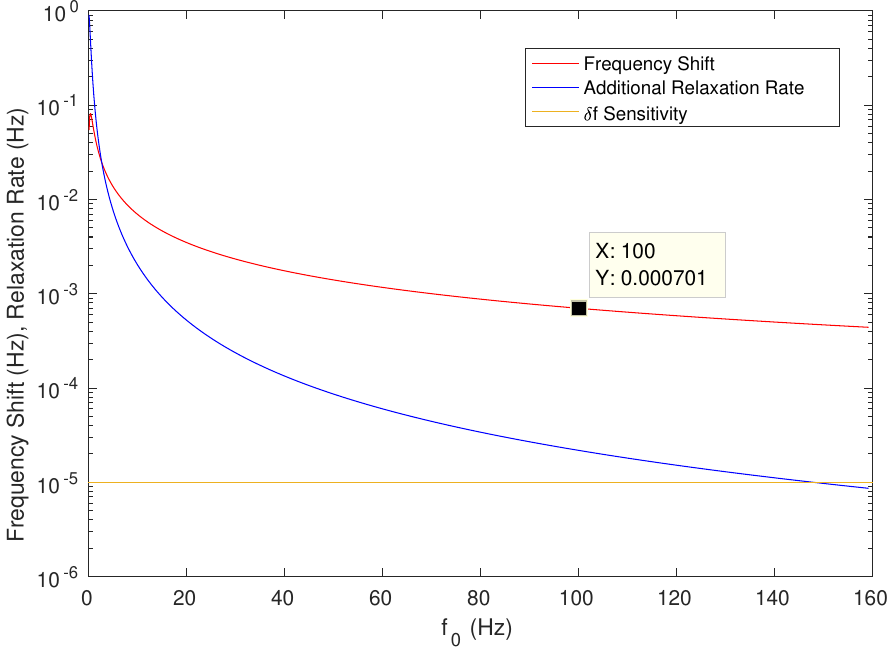}
\end{center}
\caption{The $\delta\omega_{B^2}$ frequency for $^3$He in a $10{\rm \,\mu G\,cm^{-1}}$ gradient along the $x$-direction (40~cm cell axis). The frequency sensitivity $\delta$f that is attainable with the SQUIDs in the SOS apparatus is also shown. $f_0 \approx 100{\,\rm Hz}$, corresponding to $B_0 \approx 30\,{\rm mG}$, is the operating frequency of the SOS apparatus and the nEDM@SNS experiment. For comparison, the additional contribution to the longitudinal relaxation of $^3$He ($T_{\rm 1,^3He,gradient}$) due to the same gradient is also shown.}
\label{fig:heliumBsquared}
\end{figure}

Another technique we can consider is measuring the ``$B^2$ precession frequency shift'' from gradients $G_x$ and $G_y$ that is given by:
\begin{align}
\delta \omega _{B^{2}}=\frac{\gamma ^{2}}{2}\left(G_{x}^{2}\mathrm{Im}\left[S_{xx}\left( \omega_0 \right)\right] +G_{y}^{2}\mathrm{Im}\left[S_{yy}\left(\omega_0 \right) \right]\right) .
\label{eq:B2}
\end{align}
We see that measuring $\delta\omega_{B^2}$ is attractive because it is a direct measure of the imaginary parts of $S_{ii}(\omega)$, which is required to predict $\delta_\omega$ (see Eq.~\ref{eq:dw}). To demonstrate the sensitivity of measuring $\delta\omega_{B^2}$ in the SOS apparatus, Fig.~\ref{fig:heliumBsquared} shows the $\delta\omega_{B^2}$ shift by applying $G_x = 10 \un{\mu G\,cm^{-1}}$. This is compared in the plot with the frequency sensitivity $\delta$f of the SOS apparatus's SQUID system assuming $T_{\rm 2,^3He}=400{\rm \,s}$ and a signal-to-noise ratio of $\mathrm{SNR}=40$. The size of the induced $\delta\omega_{B^2}$ is $\sim 100\times$ larger, and thus can be easily measured. This technique has the potential to be better than the $T_1$ measurements. However it has yet to be experimentally demonstrated; at the very least, it would provide a cross-check of the other techniques. Some of the systematic effects involved when determining $\delta\omega_{B^2}$ are discussed in Appendix~\ref{appen:corrFunc}.

So far we have only discussed measurements of the $^3$He correlation function. To determine the neutron correlation functions, measurements of the spin relaxation times and $\delta\omega _{B^{2}}$ can also be performed. This can be done by using the n-$^3$He capture scintillation light. Measurement of $T_{\rm 1,n}$ is discussed in Sec.~\ref{sec:cellTesting}. Measurement of $T_{\rm 2,n}$ can be done by measuring $T_{\rm 2,tot}$ from a free precession measurement. $T^{-1}_{\rm 2,tot} \approx T^{-1}_{\rm 2,n,gradient} + T^{-1}_{\rm 2,^3He,gradient}$ in the presence of sufficiently large applied field gradients, so $T_{\rm 2,n,gradient}$ can be deduced since $T_{\rm 2,^3He,gradient}$ is known from the SQUID measurements. The $\delta\omega _{B^{2}}$ shift of the neutron can also be determined with the free precession light signal, with approximate sensitivities given in Table~\ref{tab:freePrecessionUncertainties}, which is of sufficient precision.

 At this time it is not clear which of the three techniques (measuring $T_{\rm 1,gradient}$ or $\omega _{B^{2}}$ static gradients or measuring $T_{\rm 2,rf\,gradient}$ in RF gradients) is superior or what combination must be measured. Each contain different sensitivities and systematics. Further discussion of the correlation function measurements, measurement procedure, sensitivity can be found in Appendix~\ref{appen:corrFunc}.

\subsection{Simultaneous neutron and $^3$He spin manipulation \label{sec:criticalDressing}}

The nEDM@SNS experiment requires simultaneous manipulation of both the neutron and $^3$He spins to carry out the experiment. This includes the $\pi/2$ pulse which rotates the spins from parallel to perpendicular with respect to the applied $B_0$ field and the measurement technique utilizing critical spin dressing. Our sensitivity estimates to a neutron EDM of $\sim 5\times10^{-28}\,e\cdot{\rm cm}$ for the free precession mode, and $\sim 3\times10^{-28}\,e\cdot{\rm cm}$ for the dressed-spin mode that offers an independent measurement with different systematics, depend on our control of these processes. Inadequate control of our simultaneous spin manipulations will cause a worsening of the nEDM@SNS experiment from our stated sensitivities.
 
The $\pi/2$ spin rotation is, in principle, fairly straightforward. A short pulse of AC magnetic field with frequency close to the neutron precession frequency but at a fraction of the amplitude of $B_0$ is applied perpendicular to $B_0$. At the end of the pulse both the neutron and $^3$He are perpendicular to $B_0$ and begin to precess about $B_0$, but with an initial angle, or phase $\phi_0$ between the spins as described in Sec.~\ref{sec:impactParameters}. Without the ability to reproducibly set the initial phase to the level of 1 mrad (or 0.06$^\circ$) for each measurement, our nEDM sensitivity would be negatively impacted. The uncertainty in the measured frequency $\sigma_{\nu_{\rm free}}$ per free precession cycle will increase from $1.8\un{\mu Hz}$ to $2.8\un{\mu Hz}$, resulting in a corresponding worsening of our nEDM sensitivity.

The simultaneous $\pi/2$ pulse technique required to do this can be demonstrated and optimized (since there are many possible frequencies and AC B-field values that can be used) with the SOS apparatus. As shown in Table~\ref{tab:freePrecessionUncertainties}, the $1\sigma$ fitted uncertainty in $\phi_0$ from each free precession cycle will be as low as $0.6^\circ$. Therefore, by performing repeated measurements of $\phi_0$ and by comparing if the population standard deviation is larger than the expected from fitting statistics alone, the initial stability of $\phi_0$ can be deduced to $\sim 10\%$ from the $\sigma_{\phi_0} \approx 0.6^\circ$ from the fitting. This lets us reach the $0.06^\circ$ precision goal.
 
In addition sufficient reduction and stability of the $^3$He elevation angle after the $\pi/2$ pulse is required to reduce the systematic shift and the additional noise in the extracted $\nu_{\rm free}$ caused by the pseudomagnetic effect produced by the n-$^3$He spin-dependent scattering length difference.

If the $^3$He spin has an elevation angle of 0.3$^\circ$ from the plane perpendicular to $B_0$, the precession frequency shift from the pseudomagnetic field is $\sim 1.6\un{\mu Hz}$, around the same precision at which we can measure $\nu_{\rm free}$ per free precession cycle in the nEDM@SNS experiment. This shift will on average be cancelled when subtracting the difference between the two cells in the nEDM@SNS experiment. However, if insufficient control of the $^3$He tipping causing a difference in the elevation angle between the two cells, noise will be introduced when taking the frequency difference and thus reduce the sensitivity to a neutron EDM in the nEDM@SNS experiment. For small angles, the size of the pseudomagnetic frequency shift is proportional to the tipping angle. Therefore, the $^3$He elevation angle after tipping should be controlled to $\lesssim 0.03^\circ$ ($\lesssim 10\%$ of the fitting uncertainty of $1.8\un{\mu Hz}$ per free precession measurement). 

Fluctuations of the $^3$He tipping angle of $0.03^\circ$ produces a shift of only $0.16\un{\mu Hz}$ if $x_{\rm pol\,3} = 8\times 10^{-11}$. However, if $x_{\rm pol\,3} = 8\times 10^{-10}$ is used in the SOS apparatus, then this produces a shift of $1.6\un{\mu Hz}$. With $x_{\rm pol\,3} = 8\times 10^{-10}$, from Table~\ref{tab:freePrecessionUncertainties}, the noise from fitting statistics only $\sigma_{\nu_{\rm free}} = 50\un{\mu Hz}$ per free precession measurement. We will thus be able to determine additional contribution to the noise to $\sim$10\% of this by repeated measurements (as described for the $\phi_0$ measurements). This corresponds to $^3$He tipping angle fluctuations of $\sim 0.1^\circ$. This is somewhat larger than what's required but performing this measurement in the SOS apparatus will identify promising $^3$He tipping pulse techniques and ways to operate the electronics for the nEDM@SNS experiment.

Use of the spin dressing technique can also provide improved sensitivity to the neutron EDM by fixing the angle between the neutron and $^3$He spins compared to the free precession. This effectively uses the capture events more efficiently. Best sensitivity is achieved, with minimal additional noise from the pseudo-magnetic field effect, by modulating the neutron-$^3$He angle from e.g. $+\phi_0$ to $-\phi_0$. This introduces additional control parameters in the technique. Thus, while the spin dressing technique using this ``critical'' dressing is quite promising, the critical' dressing condition has not been studied experimentally and includes many variables the should be optimized, including dressing field, dressing frequency, modulation frequency, modulation waveform, $\phi_0$, etc.. 

Exploring this phase space experimentally will certainly take some time and may well include adjustments to the hardware (i.e. multiple cooldowns) such as additional AC shim coils, optimizing pick-up loops for AC B-field feedback, etc. Optimizing this technique could well take many months and multiple cooldowns to converge. Performing this optimization with the SOS Apparatus will almost certainly reduce the number of cooldowns as well as operation time required for the main apparatus as the critical dressing parameter space is studied.

\subsection{Cell testing \label{sec:cellTesting}}

Measurements of the $^3$He depolarization rate and UCN loss probability on the measurement cell walls form the basis of the cell testing program. We currently do not have a good estimate for the probability to produce a good cell. We also do not have a good estimate for how frequently we'll need to change cells (e.g., due to spark damage, formation of a leak, or magnetic contamination). But we do know that every bad cell identified by the SOS apparatus -- and therefore not installed -- will save a cooldown cycle.
 
\begin{figure}[htbp]
\begin{center}
\includegraphics[width = 13cm]{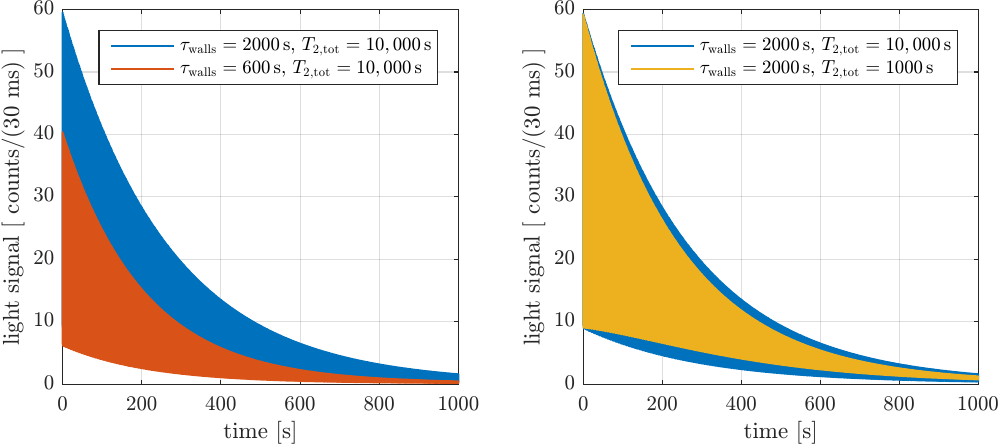}
\caption{Example of how ultracold neutron losses and depolarization of neutron and $^3$He affects the scintillation light signal from free precession measurement and reduces the statistical sensitivity on the nEDM@SNS experiment. The lines shown are those from fitting Monte Carlo generated light signals using Eq.~\ref{eq:FIDsignal}. The oscillations at $\omega_{\rm free} \approx 100\un{Hz}$ can't be resolved visually in these plots.}
\label{fig:lightSignalCell}
\end{center}
\end{figure}

Loss of ultracold neutrons and depolarization of neutrons and $^3$He cause a reduction in statistical sensitivity. Plots showing how reductions in $\tau_{\rm tot}$ and $T_{\rm 2,tot}$ affect the scintillation light signal for free precession measurements are shown in Fig.~\ref{fig:lightSignalCell}. In these examples, a reduction of $\tau_{\rm tot}$ from the nEDM@SNS design goal of 2000\un{s} to 600\un{s}, and the design goal $T_{\rm 2,tot}$ from 10,000\un{s} to 1000\un{s}, both cause a reduction in sensitivity to a neutron EDM in the final nEDM@SNS experiment by a factor of $\sim 2$. 

Effects that can cause increases in UCN loss and neutron and $^3$He depolarization include:
\begin{itemize}
\setlength{\itemsep}{1pt}
\item coating imperfections at the cell walls;
\item gluing imperfections at the cell inner corners;
\item contamination (nuclear or magnetic) inside the cell;
\item coating micro-roughness.
\end{itemize}
The current active measurement cell and UCN storage program so far has focused on reducing neutron losses only. Ultracold neutron storage measurements have been performed in vacuum down to $\sim 15\un{K}$. While studies have been done on $^3$He depolarization on small-scale coated cells ($\sim 1'' \times 1'' \times 1''$)  by our collaboration at close to our operating conditions ($\sim 0.4\un{K}$ and $B_0 \sim 7\un{G}$), they have yet to be done on full-sized cells. Neutron depolarization measurements have also not been performed on our coatings.

The SOS apparatus will be able to measure the loss and depolarization properties of full-sized measurement cells at the final operating conditions (final temperature in superfluid helium and close to final magnetic fields) in a single apparatus. Use of full-sized, final cells are important since the production process produces different coating qualities and possibility different contamination. New magnetic effects might appear at the final temperatures as well. 

It should be emphasized that after the cell properties have been verified in the SOS apparatus, the same cells will be installed in the full nEDM apparatus, significantly reducing the risk and time costs of installing a poor quality cell. Contamination on the inside of a cell is most likely to occur before (e.g. already in the coating solution) and during cell production (e.g. airborne dust or magnetic particles) due to the large surface area and exposure time. From previous UCN storage measurements, we have demonstrated that reinstalling the cell after exposure to air and repeated cooldowns does not degrade the neutron storage properties of the cell. 
 

The $\tau_{\rm tot}$ of a cell can be measured by using the exponentially decaying scintillation light from $\beta$-decay without $^3$He. A day of measurements on the SOS apparatus will produce $\sigma_{\tau_{\rm tot}} \lesssim 6{\rm \, s}$ for a UCN density $\rho_{\rm UCN} = 30{\, \rm UCN\,cm^{-3}}$, corresponding to $\sigma_{\tau_{\rm walls}} \lesssim 20{\rm \,s}$, which will be of sufficient accuracy.

The wall contributions to $T_{\rm 2,tot}$ (see Eq.~\ref{eq:tautot_tau2tot}) are given by $T_{\rm 2,n,walls} = 2 \,T_{\rm 1,n,walls}$ and $T_{\rm 2,^3He,walls} = 2 \,T_{\rm 1,^3He,walls}$. Therefore, it is sufficient to determine the longitudinal relaxation times $T_{\rm 1,walls}$. Generally, one measures $1/T_{\rm 1, tot} = 1/T_{\rm 1, walls} + 1/T_{\rm 1, gradient}$. However, in the SOS apparatus' magnetic field gradients of $\sim 500\un{nG/cm}$ the expected $T_{\rm 1, gradient} \gtrsim 10^{6}\un{sec}$ for both $^3$He and neutrons. Therefore, $T_{\rm 1,^3He, tot} \approx T_{\rm 1,^3He, walls}$ and $T_{\rm 1,n, tot} \approx T_{\rm 1,n, walls}$, which avoids the need of separating out the two longitudinal relaxation contributions.

$T_{\rm 1,^3He, walls}$ can be measured with the SQUID magnetometer using a common NMR technique where the $^3$He spin is repeated tipped by a small angle ($\sim \pi/4$) from $B_0$ with long known time delays in between (e.g. $\sim 1000\un{sec}$). Due to the large polarized $^3$He concentration that can be used for these measurements in the SOS apparatus, the desirable precision can be easily obtained.

To determine $T_{\rm 1,n,walls}$, the scintillation light signal when leaving both the neutron and $^3$He spins aligned with $B_0$ can be used. Knowledge of $\tau_{\rm tot}$ and $T_{\rm 1,^3He, walls}$ from the two previous measurements can then be used to extract $T_{\rm 1,n,walls}$. This is described in more details in Appendix~\ref{appen:T1neutronMeasure}.

\section{Experimental design differences between nEDM@SNS and SOS apparatus}

The experimental design differences between the SOS apparatus and the nEDM@SNS experiment summarized in this section are what allow the size, complexity and turnaround times of the former to be reduced relative to the latter. The SOS apparatus will have a single full-sized measurement cell filled with isotopically pure superfluid helium at the final operating temperature of $\sim 0.4\un{K}$. The cell hole sealing geometry and interface to the vestibule of the 3-position valve of the SOS is designed so that it can be used with the V1-valve system of the nEDM experiment. Therefore, the cells tested in the SOS apparatus will be fully transferrable to nEDM after testing. Indeed, this is one of the main goals of the measurement program described in Sec.~\ref{sec:cellTesting}.

There will be no high voltage system in the SOS apparatus. However, the systematic error resulting from the interaction of the motional ($\vec{E}\times \vec{v}/c$) magnetic field with stray magnetic field gradients (often called the ``geometric phase'' induced false EDM effect) can still be predicted using the SOS apparatus. As described in more details in Sec.~\ref{sec:correlationFunc}, this is done by studying the motion of the UCNs and $^3$He via characterization of the correlation functions. These studies do not require an electric field. Furthermore, a dummy ground electrode can be installed in the SOS apparatus for instrumentation tests of the SQUID system.

The ultracold neutrons used in the SOS apparatus will be loaded into the measurement cell from an external source rather than produced \emph{in-situ} with the FNPB cold neutron beam for nEDM@SNS. This results in absence of any activation background as well as a reduction in the initial UCN number of around an order of magnitude. The setup can also be upgraded to run at a cold neutron beam.

\begin{table}[tbp]
\caption{Summary of experimental design differences between the full nEDM@SNS experiment and the SOS apparatus at PULSTAR}
\begin{center}
\begin{tabular}{|c | p{5cm}| p{7cm} |}
\hline
 & \emph{nEDM@SNS} & \emph{SOS apparatus} \\ \hline \hline
measurement cell & 2$\times$ cells & 1$\times$ full-sized cell, designed so can be installed in nEDM\\ \hline
high voltage & direct-feed and then Cavallo & none\\ \hline
$0.4\,{\rm K}$ superfluid He & $\sim 1000{\,\rm L}$ & $\sim 5{\,\rm L}$ \\ \hline
ultracold neutrons & UCN production inside of cell with the FNPB cold neutron beam & UCNs fed in from external UCN source  \\ \hline
polarized $^3$He & ABS system: $x_3\sim 10^{-10}$ and $P_3\sim 98\%$) & MEOP system: $x_3$ can reach $10^{-7}$ with $P_3\sim 70\%$  \\ \hline
magnetic field gradients & $< 100\un{nG/cm}$ or  $< 3 \times 10^{-6}\un{cm^{-1}}$ for $T_{\rm gradient, ^3He} > 10,000\un{sec}$ & $<500\un{nG/cm}$ or $< 1.5 \times 10^{-5}\un{cm^{-1}}$ for $T_{\rm gradient, ^3 He} > 400\un{sec}$ \\ \hline
measurement cycle rate & 30$\times$/day (dead time = 400\un{s}, 24 hours/day) & 3$\times$/day (dead time = 2\un{hrs}, 9 hours/day) \\ \hline
\end{tabular}
\end{center}
\label{table:expDifferences}
\end{table}

A MEOP polarized $^3$He source at room temperature will be used in the SOS apparatus rather than the cryogenic atomic beam source (ABS) which will be used in the nEDM@SNS apparatus. The MEOP system compared to the ABS can produce 2-3 orders of magnitude more polarized $^3$He atoms with the trade-off of a reduced polarization ($P_3 \sim 98\% \textrm{ vs. } 70\%$). A higher number of polarized $^3$He atoms will make measurements using only $^3$He and SQUIDs more easily performed and with better statistics.

The trade-off with the MEOP system's lower $P_3$ is that achieving the same SQUID $^3$He signal strength, the n-$^3$He absorption loss will be higher. For example, in the nEDM@SNS experiment the optimal is $x_{\rm pol \,3} = 8\times 10^{-11}$ for extracting the oscillating frequency from the free precession light signal (see Sec.~\ref{sec:impactParameters}). If the noise in the SQUIDs of the SOS apparatus are the same as in the nEDM experiment, then to have the same signal-to-noise ratio the neutron-$^3$He absorption time constant $\tau_3$ will decrease from 500\un{s} to $350\un{s}$ (see Eq.~\ref{eq:IFtau3}). This shortens the useful observation time with ultracold neutrons in the SOS apparatus.

The magnetic field homogeneity that can be achieved in the SOS apparatus is reduced due to the reduced available space and size compared to nEDM@SNS. However, the magnetic field requirements in the SOS apparatus are reduced because of the shorter observation time caused by the reduced $P_3$ discussed above. The transverse relaxation due to magnetic field gradients $T_{\rm 2, gradients}$ is proportional to the square of the linear gradients in the cell (e.g. see Eq.~\ref{eq:dw}). The specified $T_{\rm 2, gradients}$ required in the SOS apparatus decreases to $\sim 400\un{sec}$ compared with $\sim 10,000\un{sec}$ in SNS@nEDM (e.g. $\sim 5 \times$ poorer gradients).

An overview of these design differences is found in Table~\ref{table:expDifferences}. In summary, 
the SOS apparatus operates in a different part of experimental parameter phase space, with less required field uniformity and lower $^3$He polarization (but with the possibility of much higher concentrations). 

\vspace{1cm}

\begin{appendices}
\addtocontents{toc}{\protect\setcounter{tocdepth}{0}}
\renewcommand{\setthesection}{\arabic{section}}
\renewcommand{\setthesubsection}{\arabic{subsection}}
\renewcommand{\thesection}{\arabic{section}}
\renewcommand{\thesubsection}{\Alph{subsection}}

\section{Uncertainties of experimental extractable parameters \label{sec:impactParameters}}

To aid the discussion of the SOS measurement program and to show that it can be successfully achieved, a more quantitative discussion in terms of the uncertainties of experimental extractable parameters from the free precession mode light signal is described.

To illustrate the impact on the parameter uncertainties for the difference in designs, the light signal expected from the free precession measurements is simulated. For this case the number of detected scintillation light events $y_i$ in a time bin with width $\Delta t$ at time $t_i$ during a free precession mode measurement is given by:
\begin{equation}
\frac{y(t_i)}{\Delta t} = I_0 {\rm e}^{-t_i/\tau_{\rm tot}} \left[1-F {\rm e}^{-t_i/ T_{\rm 2,tot}} \cos(\omega_{\rm free} t_i + \phi_0) \right] + R_{\rm BG} \;,
\label{eq:FIDsignal}
\end{equation}
where
\begin{equation}
I_0 = N_0 \left( \frac{\epsilon_\beta}{\tau_\beta}+\frac{\epsilon_3}{\tau_3} \right) \quad \textrm{,} \quad F = \frac{\epsilon_3 P_3 P_n}{\tau_3 \left(\frac{\epsilon_\beta}{\tau_\beta}+ \frac{\epsilon_3}{\tau_3} \right)} \quad \textrm{and} \quad \tau_{\rm 3} = \frac{2}{n\,v_\sigma \, \sigma_{\rm \uparrow\downarrow}(v_\sigma)\,  x_{3}} \;.
\label{eq:IFtau3}
\end{equation}
$N_0$ is the initial number of neutrons in the cell, $\epsilon_\beta$ is the detection efficiency for neutron $\beta$-decay (${\rm n \rightarrow p + e + \overline{\nu}_e}$) events, $\tau_\beta$ is the neutron $\beta$-decay mean lifetime of $\sim 880\un{s}$, $\epsilon_3$ is the detection efficiency of a n-$^3$He capture event, $\tau_3 $ is the time-averaged n-$^3$He capture rate determined by the amount of polarized $^3$He loaded into the cell ($n$ is the helium number density, $\sigma_{\rm \uparrow\downarrow}(v_\sigma)$ is the cross-section of n-$^3$He capture in the singlet state at the velocity $v_\sigma$, which for $v_\sigma=2200{\rm\,m\,s^{-1}}$, $\sigma_{\rm \uparrow\downarrow} = 11{\rm \, kb}$), $P_3$ and $P_n$ are the initial $^3$He and neutron polarizations, $\omega_{\rm free}$ is the angular beating frequency between the $^3$He and neutron spins ($\omega_{\rm free}  =2\pi\nu_{\rm free}= \gamma_3 B_0-\gamma_n B_0$), $\phi_0$ is the initial phase, and $R_{\rm BG}$ is the ambient background rate.

\begin{figure}[tbp]
\begin{center}
\includegraphics[width = 7cm]{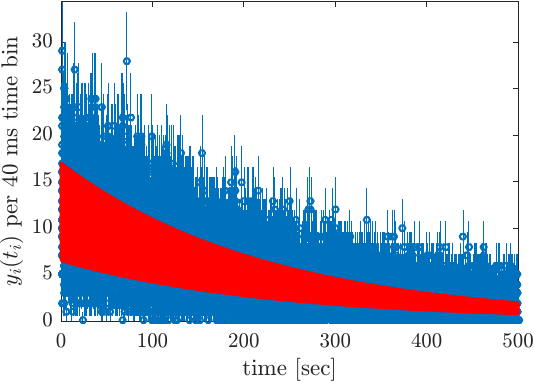}
\includegraphics[width = 7cm]{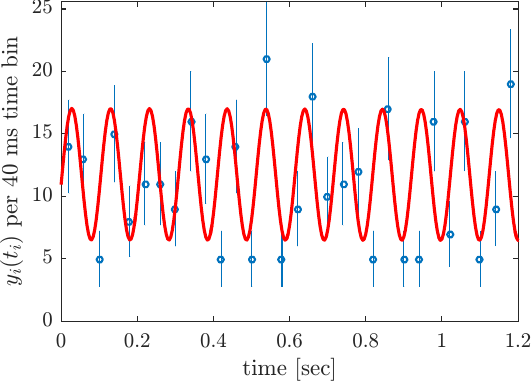}
\caption{The simulated scintillation light signal expected from the Systematics and Operational Studies apparatus during a free precession measurement. The red solid line is a fit to the binned data points. Shown on the right is a zoomed in view of the plot on the left.}
\label{fig:freePrecessionSignal}
\end{center}
\end{figure}

The contribution from losses and depolarization enter into Eq.~\ref{eq:FIDsignal} through the total neutron storage time $\tau_{\rm tot}$ and the total transverse polarization relaxation time $T_{\rm 2,tot}$ given by:
\begin{equation}
\label{eq:tautot_tau2tot}
\frac{1}{\tau_{\rm tot}} = \frac{1}{\tau_3} + \frac{1}{\tau_\beta} + \frac{1}{\tau_{\rm wall}} \quad \textrm{and}\quad
\frac{1}{T_{\rm 2,tot}} = \frac{1}{T_{\rm 2,n,wall}} + \frac{1}{T_{\rm 2,n,gradient}} + \frac{1}{T_{\rm 2,^3He,wall}} + \frac{1}{T_{\rm 2,^3He,gradient}} \;,
\end{equation}
where the ``wall'' subscripts are for contributions from the walls of the cell and the ``gradient'' subscripts in the expression for Eq.~(\ref{eq:tautot_tau2tot}) refer to depolarization caused by magnetic field gradients. The wall contributions will be the same between SOS and nEDM (since the same cell that will be used) whereas the gradient contributions will be different (because of the different magnetic field coil and shielding).


The simulated light signal from a free precession measurement in the SOS apparatus is shown in Fig.~\ref{fig:freePrecessionSignal}. From fitting this data using an iterative re-weighted least squares fitting algorithm, the $1\sigma$ uncertainty in extracting the key parameters, $\sigma_{\tau_{\rm tot}}$, $\sigma_{\nu_{\rm free}}$ and $\sigma_{\phi_0}$ (leaving all these parameters free in the fit), are shown in Table~\ref{tab:freePrecessionUncertainties}. Note that these uncertainties assume that $\phi_0$ is left as a free parameter in the fit (see Sec.~\ref{sec:criticalDressing} for a discussion of reducing this uncertainty).

\begin{table}[h]
\caption{Uncertainties in the extracted parameters in the SOS apparatus per free precession cycle assuming: $P_{\rm n} = 0.98$, $\tau_{\rm walls} = 2000{\rm \,s}$, $\epsilon_3 = 0.93$, $\epsilon_\beta = 0.50$, $T_{2,\rm tot} = 400\un{sec}$, and $R_{\rm BG} = 5\un{Hz}$. The UCN density $\rho_{\rm UCN}$ and $^3$He polarization $P_3$ are varied to see its impact. The effects of deliberately increasing $x_{\rm pol\,3}$, which is possible due to the MEOP source, are also shown.} 
\small
\begin{tabular}{c c c c c c c c} \hline
$T_{\rm 2,\,tot}$ & $\rho_{\rm UCN}$ & $P_3$ &$ x_{\rm pol\,3}$ & $\tau_3$ & $\sigma_{\tau_{\rm tot}}$ & $\sigma_{\nu_{\rm free}}$ & $\sigma_{\phi_0}$ \\ \hline \hline
\multirow{8}{*}{400 s} & \multirow{4}{*}{10\un{cm^{-3}}} & \multirow{2}{*}{0.5} & $8\times10^{-11}$ & 250\un{s} & 2\un{s} & $60\un{\mu Hz}$ &  2.5$^\circ$ \\ \cline{4-8}
& & & $8\times10^{-10}$ &  25\un{s} & 0.2\un{s} & $150\un{\mu Hz}$ &  1.6$^\circ$ \\ \cline{3-8}
& & \multirow{2}{*}{0.7} & $8\times10^{-11}$ & 350\un{s} & 3\un{s} & $40\un{\mu Hz}$ &  2.1$^\circ$ \\ \cline{4-8}
& & & $8\times10^{-10}$ &  35\un{s} & 0.2\un{s} & $80\un{\mu Hz}$ &  1.1$^\circ$ \\ \cline{2-8}
& \multirow{4}{*}{30\un{cm^{-3}}} & \multirow{2}{*}{0.5} & $8\times10^{-11}$ & 250\un{s} & 1\un{s} & $30\un{\mu Hz}$ &  1.4$^\circ$ \\ \cline{4-8}
& & & $8\times10^{-10}$ & 25\un{s} & 0.1\un{s} & $80\un{\mu Hz}$ &  0.9$^\circ$ \\ \cline{3-8}
& & \multirow{3}{*}{0.7} & $8\times10^{-11}$ & 350\un{s} & 1\un{s} & $20\un{\mu Hz}$ &  1.1$^\circ$ \\ \cline{4-8}
& & & $8\times10^{-10}$  & 35\un{s} & 0.1\un{s} & $50\un{\mu Hz}$ &  0.6$^\circ$ \\ \cline{4-8}
& & & $8\times10^{-9}$  & 3.5\un{s} & 0.03\un{s} & $450\un{\mu Hz}$ &  0.6$^\circ$ \\ \cline{1-8}
\end{tabular}
\label{tab:freePrecessionUncertainties}
\end{table}

The polarized $^3$He concentration is scanned to illustrate the usefulness of having access to a wide range of $x_{\rm pol\,3}$. In measurements where extraction of the neutron precession frequency is important, a low $x_{\rm pol\,3} \sim 10^{-10}$ will be most useful. When extraction of the initial phase $\phi_0$ is important, an intermediate $x_{\rm pol\,3} \sim 10^{-9}$ will be most useful. Finally, when access to the polarized $^3$He precession frequency or signal strength needs to be enhanced, operation with $x_{\rm pol\,3} \sim 10^{-8} \textrm{ to } 10^{-7}$ will be most useful. For $x_{\rm pol\,3} \lesssim 10^{-7}$ the mean-free-path of the $^3$He remains essentially unchanged since it is still in the regime where scattering with phonons in the superfluid helium dominates.


\section{Detailed description of apparatus \label{appen:overviewApparatus}}

\begin{figure}[htbp]
\begin{center}
\includegraphics[width=10cm]{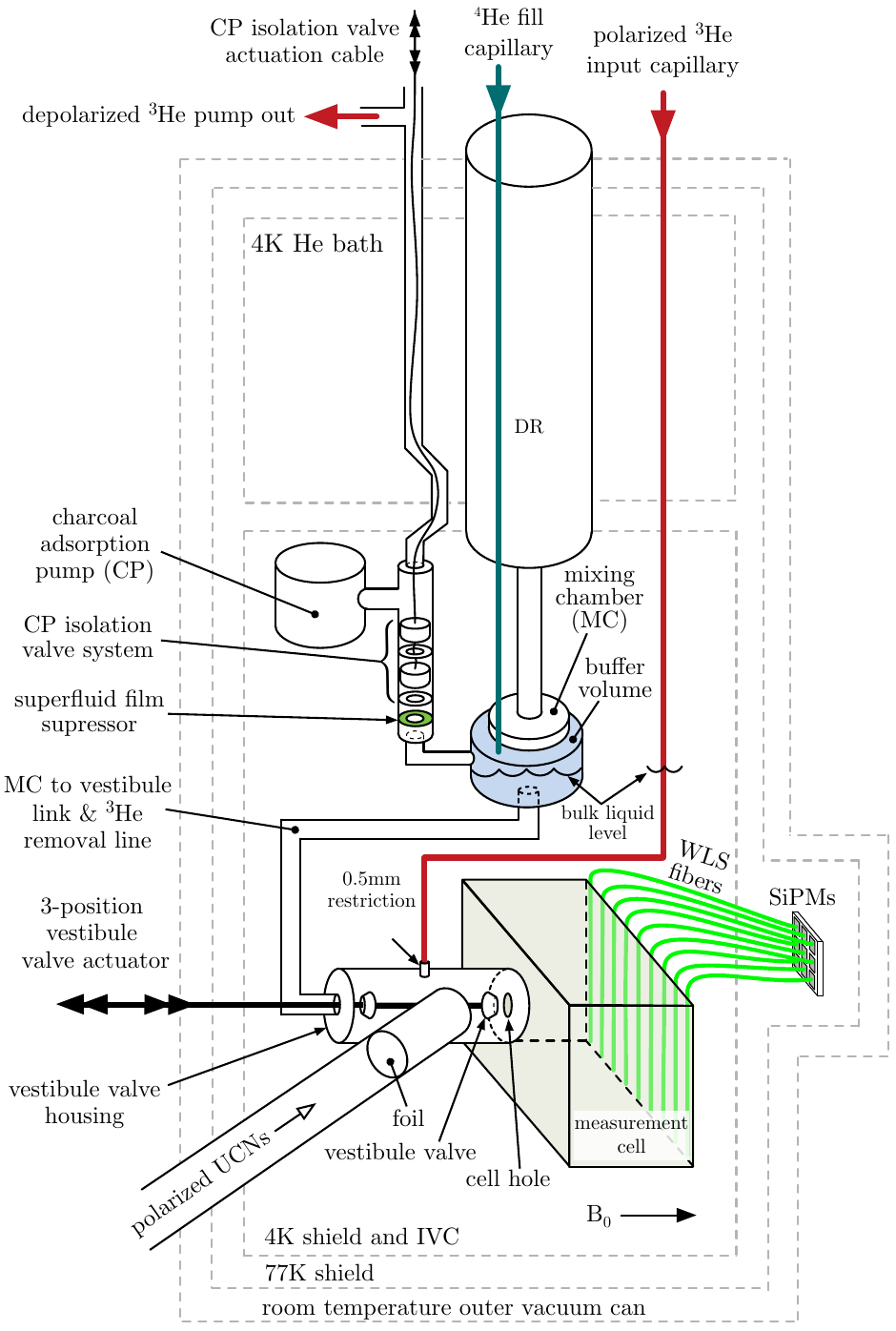}
\caption{Schematic of the SOS apparatus in Phase II. The magnetic coils, SQUID magnetometers and pick-up loops are not shown and many components are shown simplified. CAD drawings of the apparatus are in Figs.~\ref{fig:CAD_verticalAll} and \ref{fig:CAD_horizontalLower}.}
\label{fig:PULSTARapparatusSchematic}
\end{center}
\end{figure}

A schematic of the experimental setup is shown in Fig.~\ref{fig:PULSTARapparatusSchematic}. Isotopically pure $^4$He is inputted and condensed via the $^4$He fill capillary to fill the $\sim 5\un{L}$ total volume of: the measurement cell; the ``vestibule'', our name for the housing of a 3-position valve system; the ``buffer volume'', a $\sim 5\un{in}$ diameter volume thermally connected to the mixing chamber (MC) of the dilution refrigeration (DR) using sinter as heat exchanger.  

The vestibule connects the cell hole to the polarized $^3$He input capillary, the ultracold neutron (UCN) guide, and the $^3$He removal line (which also acts as the thermal link with the MC). With the 3-position vestibule valve in the left-hand most position in Fig.~\ref{fig:PULSTARapparatusSchematic}, polarized UCNs and polarized $^3$He can be loaded into the measurement cell. Then moving the vestibule valve to the right-hand most position, the UCNs and $^3$He are confined in the cell for the measurements. The inner walls and the valve piston will be coated with deuterated plastic for good UCN reflection and low $^3$He depolarization losses. The measurement cell will also be coated with deuterated plastic but with a deuterated fluorescent dye added for converting the superfluid helium scintillation light peaked at 80\un{nm} to the blue region. At the end of a measurement, the vestibule valve is moved to the center position to allow the depolarized $^3$He to be removed via the $^3$He removal line.

\begin{figure}[htbp]
\begin{center}
\includegraphics[width=10cm]{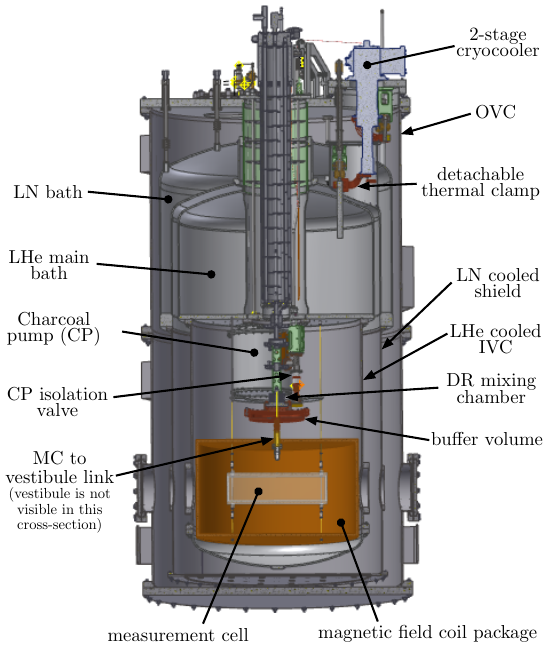}
\caption{3D model of the apparatus. Shown as a vertical slice coincident on long central axis of the measurement cell. The WLS fibers, SiPMs and MEOP $^3$He system are not shown here. The magnetic field coil package is shown schematically only.}
\label{fig:CAD_verticalAll}
\end{center}
\end{figure}

Inside the inner vacuum chamber (IVC) below the height of the buffer volume, the components will be made from only plastics and composite materials to avoid magnetic inhomogeneities, as well as reducing RF-induced eddy current heating. The outer walls and thermal shields of the cryostat are made with aluminum and other non-magnetic materials as much as possible also. A 2-stage Gifford McMahon cryocooler (see Fig.~\ref{fig:CAD_verticalAll}) will be used for initial cooldown and to reduce cryogen consumption, but can be turned off during measurement to reduce magnetic and vibrational noise.

\begin{figure}[htbp]
\begin{center}
\includegraphics[width=10cm]{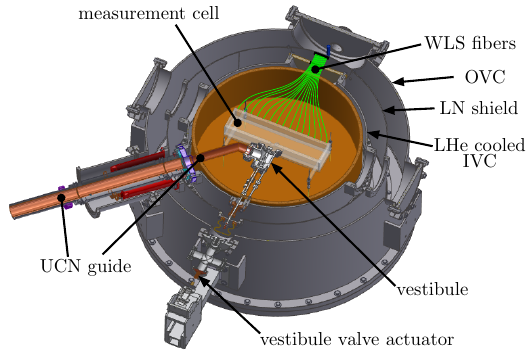}
\caption{3D model of the apparatus. Shown as a horizontal slice coincident on the long central axis of the measurement cell. The WLS, SiPMs and magnetic field coil package are shown schematically only.}
\label{fig:CAD_horizontalLower}
\end{center}
\end{figure}

The UCNs from the external source will be polarized outside of the SOS apparatus. Our apparatus is envisioned to be used at the PULSTAR UCN source, which has been calculated to provide a density of  $30\un{UCN\,cm^{-3}}$ (unpolarized) at the source exit for the reactor operating at 1\un{MW}, with there being a planned upgrade to 2\un{MW} . To maintain neutron polarization non-depolarizing coatings of the UCN guides will be used. Inside the apparatus, the guide will transition from room temperature to the vestibule (see Fig.~\ref{fig:CAD_horizontalLower}).  There will be a a thin-foil 4K window and a 45$^\circ$ bend section just before the vestibule to reduce radiation heating. There will be a thin-foil at the UCN guide entrance to the vestibule for containing the superfluid helium. Biaxially oriented polypropylene 0.0013\un{in} thick has been tested to be superfluid tight and can withstand more than 1\un{bar} pressure differential when cold. This thin foil with its near zero neutron optical potential will have a low UCN transmission loss.
 
The source of polarized $^3$He will be a Metastable Exchange Optical Pumping (MEOP) system located above the cryostat connected to the polarized $^3$He input capillary.  A key difference between the SOS apparatus is the higher density of polarized $^3$He that can be achieved in the cell compared with the atomic beam source (ABS) of the SNS nEDM experiment. The latter is optimized for $x_{\rm pol\,3} \sim 10^{-10}$, equivalent to $\sim 10^{12}\un{cm^{-3}}$ density in the cell. The SOS apparatus' MEOP system will produce $\sim 70\%$ polarized $^3$He of $\sim 1\un{Torr}$ gas at room temperature ($\sim 3 \times 10^{16}\un{cm^{-3}}$ density) in a volume $\sim 5\%$ relative to the superfluid volume in the SOS apparatus. Therefore, even after injection of the $^3$He, many orders of magnitude higher polarized $^3$He density can be loaded into the cell compared to the ABS system. A pre-holding volume will be used to control how much $^3$He is injected. Of course, the trade-off is a lower polarization. A thin 2\un{mm} tube made from glass and kapton thermally anchored at appropriate places and passing through the 4\un{K} LHe bath will be used to reduce the heat load to the system during injection.

After a measurement, the $^3$He from the cell will be removed via the $^3$He removal line. When the vestibule valve is open the $^3$He in the cell will diffuse up to the buffer volume until the $^3$He density is constant. The liquid surface in the buffer volume will serve as an evaporator. The buffer volume is connected to a charcoal adsorption pump (CP), with a superfluid film suppression system (e.g. a film burner and film pinner), so that mostly gaseous helium from the buffer volume is pumped by the CP. The CP is a sealed volume containing $\sim 1 \un{kg}$ of activated charcoal that is cooled with a thermal link to the 4~K LHe bath. 

In equilibrium, the ratio of the $^3$He density in the vapor just above the liquid in the buffer volume and the density of the $^3$He dissolved in the superfluid just below the liquid level is given by: $n_{\rm {^3}He\,gas}/n_{\rm ^3He\,liquid}\approx 2.4^{-3/2}{\rm e}^{-(2.8\un{K})/T}$, where $n_{\rm {}^3{He}\,liquid} =  x_3 \times (2.2\times10^{22}\un{cm^{-3}})$. And the partial pressure of $^3$He in the gas is given by $P_{\rm ^3He} = n_{\rm ^3He\,gas} k_B T$. 

The removal rate of $^3$He is $\dot{Q}_{\rm ^3He} = P_{\rm ^3He} S$, where $S$ is the pumping speed of the combined CP and conductance of the piping connecting the buffer volume to the CP. For our geometry, the pumping speed is dominated by the restriction of the piping. Therefore, if the temperature of the evaporator is increased from our nominal $0.4\un{K}$, $\dot{Q}_{\rm ^3He}$ will also increase; for example from $0.4\un{K}$ to $0.5\un{K}$, $\dot{Q}_{\rm ^3He}$ increases by a factor of $\sim 5$. This neglects the further increase due to the increase in conductance in the molecular regime $C\propto T^{1/2}$. 


The capacity of the CP for helium at 4.2\un{K}, assuming a pressure of $10^{-7}\un{mbar}$, is $10\un{mol\,kg^{-1}}$. The amount of $^3$He in the system if $x_3 \sim 10^{-10}$ is used is $\sim10^{-8}\un{mol}$. Thus, the CP has enormous capacity compared simply the amount of $^3$He required to be removed. However, a much larger amount of evaporated $^4$He will get adsorbed during the $^3$He removal cycle. The partial pressure of $^4$He, given by the vapor pressure, is $P_{\rm ^4He} \approx (680\un{mbar})\times {\rm e}^{-(8.6\un{K})/T}$, so the ratio of the partial pressures:
\begin{equation}
\frac{P_{\rm ^4He\, gas}}{P_{\rm ^3He\, gas}} \approx \frac{ (0.35\un{K})}{x_3 T} {\rm e}^{-(6.1\un{K})/T}
\end{equation}
Therefore, if a higher $T$ is used in the evaporator, more $^4$He will be evaporated and be adsorbed by the CP. The optimal operating conditions will be depend on balancing between these effects. To further reduce $^4$He from saturating the CP pump, a superfluid film suppressor and pinner will be used. 
 
If the temperature in the evaporator needs to be raised during evaporation, it will be advantageous to raise the temperature by applying a heat load ($\sim 1-2\un{mW}$) using a heater at the measurement cell. By allowing  phonons to flow from the measurement cell to the buffer volume, an added heat flush effect will be applied on the $^3$He and increase the density of $^3$He in the buffer volume, thus further speeding up the $^3$He removal time. With a buffer volume of $\sim 0.2\un{L}$, the $x_3$ at the evaporator will be increased by $\gtrsim 10$ times, reducing the required evaporation time by a similar factor. The $^3$He removal process is expected to take $\lesssim$ 3 hours after, which the superfluid helium will be cooled back down to the nominal $0.4\un{K}$ in preparation for the next measurement.

The $^3$He removal system is designed so that it can be purged while the cryostat remains cold. For this procedure, the CP isolation valve is closed and the CP pump heated up with the thermal link between it and the 4~K LHe bath switched off. The desorbed $^3$He and $^4$He is then pumped out to room temperature. Due to the long pipe and small diameter, this process is expected to take a day or so to completely purge the charcoal.

During measurement, signals from the spin-dependent ${\rm n}+{}^3{\rm He}$ capture events provides information on the neutron spin. After a ${\rm n}+{}^3{\rm He} \rightarrow {\rm p} + {}^3{\rm H}$ event, the 764\un{keV} of kinetic energy will produce $\sim 6,400$ EUV scintillation photons in the superfluid helium peaked at 80\un{nm}. These EUV photons will be down-converted to blue photons by the deuterated tetraphenyl butadiene (dTPB) doped dPS coating on the inner surface of the measurement cell. These blue photons are further down-converted into green photons at the core of 1.5\un{mm} diameter wavelength shifting (WLS) fibers coupled to the outer walls of the cell. These fibers guide out the green photons to silicon photomultipliers (SiPMs) detectors situated in a volume at $\sim 100\un{K}$, with an optical break at the IVC. The fibers will be coupled to one wall of the cell. There will also be dielectric reflecting film on the outside of the fibers and around the cell (not shown in the figures) to increase the light capture efficiency. An estimate of $\sim 30$ photo-electrons (PEs) will be produced by the SiPMs, taking into account the photon conversion efficiencies, capture and guiding efficiencies, transmission losses, and quantum efficiency of the SiPMs.

The magnetic coils, shown in Figs.~\ref{fig:CAD_verticalAll} and \ref{fig:CAD_horizontalLower}, will be wires fixed onto a shell in the form of cosine $\theta$ coils. These coils provide the $B_0$ (horizontal and perpendicular to the long-axis of the cell), and RF-fields for spin-flipping and spin-dressing (horizontal and parallel to the long axis of the cell). Sheets of superconductors (not shown) will be used as a magnetic shield as well as shaping the flux return in order to achieve a high field uniformity.

To minimize operational costs of the cryostat there is an embedded cryocooler thermally linked to the main LHe bath, which significantly reduces LHe consumption. In addition,  a 40 liters per day capacity He reliquefier was added to the system to recycle evaporated LHe. The recycled He is collected in the 250l storage Dewar, which in turn is connected to the cryostat by a custom designed transfer tube with a LN$_2$ precooling capability. As result, using LN$_2$ and cryocooler precooling  we are able to reach about 15K temperature in the main bath and then start transfer LHe slowly from the storage Dewar minimizing need to replenish our helium inventory.

\section{Measuring the correlation function \label{appen:corrFunc}}

\subsection{Measuring the correlation function of $^{3}$He}

The correlation function of $^{3}$He, must be characterized prior to measuring the correlation function of the neutrons, as the signal from the neutrons is proportional to cosine of the angle from the $^{3}$He. Therefore we only know the neutron's phase relative to the $^{3}$He. The noise in the $^{3}$He is due to noise in the SQUID, we will assume it is white noise. 

\subsubsection{Static Gradient method, T$_{1}$}

There are three ways to access the correlation function through relaxation and frequency shifts. One of these is to measure the longitudinal relaxation with a constant gradient. After application of a constant gradient in the $z$ direction such that $G_{z}=-2Gx=-2Gy$, we would measure the correlation function by the longitudinal relaxation rate due to the applied gradient. The contribution to the relaxation is given in \cite{redfield,mcgregor,pignol2015}. 

Measuring the correlation function through the longitudinal relaxation is achieved with free induction decay (FID) decay and SQUIDs in the SOS apparatus. For this measurement we apply a known $\alpha$ pulse where $\alpha \ll \frac{\pi }{2}$. This forces the $^{3}$He into a FID, which is measured by the SQUIDs. Once the decay ($T_2$*) is below the noise we apply the relaxing gradient and wait for a given time. After the time has elapsed we supply another $\alpha $ pulse, measure the FID, and repeat. This measurement procedure continues for 5 to 6 $T_{1}$ time constants so the background is well known. 

This measurement has a set of competing relaxation processes that are considered systematic effects, these effects can largely be corrected for with calibration measurements, within the precision of the corrections. The total longitudinal relaxation rate is given by the formula:
\begin{align}
\frac{1}{T_{\rm 1,tot}}=\frac{1}{T_{\rm 1,wall}}+\frac{1}{T_{\rm 1,gradient}}+\frac{1}{T_{\rm pulse}} \;,
\end{align}
where $T_{\rm 1,wall}$ is the relaxation for the cell walls, $T_{\text{pulse}}$ is the loss from the $\alpha $ pulse, and $T_{\rm 1,gradient}$ is the quantity we are trying to measure. The precision of the measurement is given by:
\begin{align}
\left( \frac{\sigma _{\text{gradient}}}{T_{\rm 1,gradient}}\right)^{2}=\left( \frac{\sigma _{\text{measured}}}{T_{\rm 1,tot}}\right) ^{2}+\left( \frac{\sigma _{\text{wall}}}{T_{\rm 1,wall}}\right) ^{2}+\left( \frac{\sigma _{\text{pulse}}}{T_{\text{pulse}}}\right) ^{2}
\end{align}

Typically $T_{\rm 1,gradient}$ can be made much shorter than $T_{\rm 1,wall}$, and the error on the pulse loss and the wall relaxation can be made very small so the only contribution is the error of the gradient measurement. If this is not the case, the density of $^{3}$He can be increased for a better calibration of the undesired relaxation, increasing density will increase signal without affecting the wall relaxation rate or the $\alpha$ pulse. 

\begin{figure}
\begin{center}
\includegraphics[width=.7\textwidth]{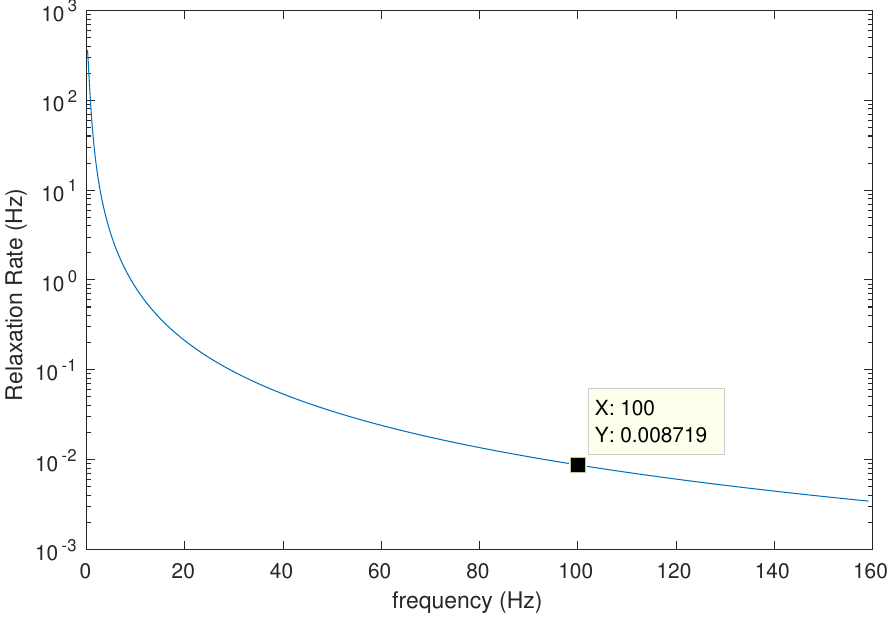}
\end{center}
\caption{Plot of the longitudinal relaxation rate (or transverse with an RF gradient) rate of $^3$He in a $200\,{\rm \mu G\,cm^{-1}}$ gradient along the 40 cm cell direction.}
\label{fig:T2relaxation}
\end{figure}

An appropriate choice for the gradient is $200{\,\rm \mu G\,cm^{-1}}$, and the relaxation rate as a function of holding field is shown in Fig.~\ref{fig:T2relaxation}. 

\subsubsection{RF Gradient method, T$_{2}$}

For the cell geometry $T_2$ relaxation in an RF gradient field:
\begin{align}
\frac{1}{T_{\rm 2,rf\,gradient}}=\frac{\gamma ^{2}G_{zx}^{2}}{2}\mathrm{Re}\left[S_{xx}(\omega _{rf})\right].
\end{align}
Where $G_{zx}$ represents a gradient in the $z$ field along the $x$ direction. We can account for other relaxation processes by the formula, 
\begin{align}
\frac{1}{T_{\rm 2,tot}}=\frac{1}{T_{\rm 2, wall}}+\frac{1}{T_{\rm 2,rf\,gradient}} \;.
\end{align}
Notice the absence of the $T_{\rm pulse}$ term, this is because we apply a $\frac{\pi }{2}$ pulse at the start of the measurement. The $\frac{\pi }{2}$ pulse must be well known to provide maximum signal strength, but it no longer requires a correction to the measurement. To measure $T_{\rm 2,tot}$ we first apply the $\frac{\pi }{2}$ pulse, record the FID\ decay with the SQUIDs for a given time interval long enough to give a good SNR. When this measurement period is over we turn off the SQUIDs, apply an RF gradient for a given time period, turn off the RF gradient, measure the signal, and repeat until for 5-6 $T_{\rm 2,tot}$ time constants for a good background measurement. The errors propagate as, 
\begin{align}
\left( \frac{\sigma _{\rm rf\,gradient}}{T_{\rm 2,rf\,gradient}}\right)^{2}=\left( \frac{\sigma _{\text{measured}}}{T_{\rm 2,tot}}\right) ^{2}+\left( \frac{\sigma _{\text{wall}}}{T_{\rm 2,wall}}\right) ^{2}.
\end{align}
because there is no systematic effect from errors in the pulse calibration
this method is preferred over $T_1$, however it may take effort in tuning the RF gradient. 

\subsubsection{Measuring the B$^{2}$ shift}

A very promising option is to measure the B$^{2}$ shift. Given by the formula \cite{pignol2015},
\begin{align}
\delta \omega _{B^{2}}=\frac{\gamma ^{2}}{2}\left(
G_{x}^{2}\mathrm{Im}\left[S_{xx}\left( \omega \right)\right] +G_{y}^{2} \mathrm{Im}\left[S_{yy}\left(
\omega \right) \right] \right).
\end{align}

This is attractive because it is a  direct measure of the imaginary component of the correlation function. While it may be challenging, in principle we can apply a gradient to excite the shift (shift gradient) along $B_x$ in $x$ and an opposing gradient in $B_y$ in y, leaving a uniform field in the $B_z$ so that $T_2$ remains long. The shift will be dominated by the gradient in the cell's long dimension. When the shift gradients are energized the holding field will change slightly due to offsets in shift coils and the cell position, it will appear as a frequency shift. This can be accounted for by measuring the frequency with the shift gradient turned off (measuring $\omega_0$) and measurements with the shift gradient reversed ($\omega_{\pm I}$). The $B^2$ shift can be determined by,  
\begin{align}
\delta \omega_{B^2}= \frac{\omega_{+I}+ \omega_{-I}}{2}-\omega_0.
\end{align}
The sensitivity of the measurement is determined from the signal to noise ratio ($\mathrm{SNR}$) at the maximum in the Fourier transform, $T_2$, and the measurement time $T$, written in terms of $T_2$, $T=\alpha T_2$.
\begin{align}
\sigma_{\delta \omega_{B^2}}=\frac{1}{\mathrm{SNR}}\frac{1}{T_2}\frac{1}{1-\left(\alpha +1 \right)e^{-\alpha}} \label{eq:B2sens}
\end{align} 
With $\alpha=5$ we achieve $96\%$ of the total sensitivity. This implies a measurement time of $T\approx5000~s,$ and translates to a frequency resolution of $\Delta f \approx 2\times 10^{-4}$~Hz. However, according to the sampling theorem, when T is chosen such that the remainder of the signal is negligible, the frequency resolution can be made arbitrarily small, without distortion, by zero padding the time discrete Fourier transform, with time samples short enough to avoid aliasing.

To estimate the sensitivity in frequency if we assume a $T_2\approx400$~s and the time signal, $S(t)$ at the start of the measurement is one tenth the RMS noise then the signal to noise ratio,
 \begin{align}
\mathrm{SNR}= T_2 \frac{S(0)}{\sigma_S}=40,
\end{align}
and we have a sensitivity of $\sigma_{\delta \omega_{B^2}}\approx 6.25\times10^{-5}~\frac{\mathrm{rad}}{s}$ or $\sigma_{\delta f_{B^2}}\approx9.94\times10^{-6}~\mathrm{Hz}.$ In this scenario if a shift gradient of $10 \frac{\mu \mathrm{G}}{cm}$ is applied, the imaginary part of the correlation function could be determined to around 1\% in a single measurement. This is shown in Fig.~\ref{fig:heliumBsquared}.

\subsection{Measuring the correlation function of neutrons}

Because the neutron density is on the order of $10^{-10}$ smaller than $^3$He they cannot be read out by the SQUIDs. The spin dependent interaction with $^{3}$He will allow measurements of the neutron correlation function. The neutron correlation function can be measured in a similar way to the $^{3}$He by measuring the relaxation in a known gradient. Here we go over the implications of both the $T_{1}$, $T_{2}~$and $\delta \omega _{B^{2}}$ methods. 

\subsubsection{Longitudinal Relaxation (T$_{1}$) with static gradient}

Similar to $^{3}$He, measurement of the longitudinal relaxation of the UCNs
can be accomplished by repeatedly tipping the spins by a known small angle $\alpha $, with a given time interval between FID signals to measure the decay. The gradient must be turned off during FID to allow maximum T$_{2}$ and therefore signal strength. The overall decay (measured decay) of the polarization can be written as a sum of rates, 
\begin{align}
\frac{1}{T_{1}}=\frac{1}{\tau _{\text{bottle}}}+\frac{1}{T_{\rm 1,gradient}}+\frac{1}{T_{\text{pulse}}}+\frac{1}{\tau_3}+\frac{1}{T_{\rm 1,^3He}}
\end{align}
Where $T_{1\text{gradient}}$ is the time constant of interest, and all other terms are systematic errors. Where $\tau _{\text{bottle}}$ is a combination of the neutron lifetime in the cell and the neutron wall depolarization time, $T_{\text{pulse}}$ is the loss from tipping pulses required to measure the FID signal, and $\tau_3$ is the rate the neutrons are lost during the measurement from the spin dependent cross section, $T_{\rm 1,^3He}$ is the relaxation of the $^{3}$He. Good precision and will be needed to minimize the error of the $T_{\rm 1,gradient}$ and it will need several dedicated measurements to account for systematic errors on the measurement of $T_{\rm 1,gradient}$, presumably $T_{\rm 1,^3He}$ would already be known fairly well due to measurements of the $^{3}$He correlation function. Furthermore it can be made such that $T_{\rm 1,gradient}$ is much shorter than the other time constants, thus reducing the impact of systematic errors, except $T_{\text{pulse}}$. Errors of this relaxation process propagate like
\begin{align}
\left( \frac{\sigma _{\text{gradient}}}{T_{\rm 1,gradient}}\right)
^{2}=\left( \frac{\sigma _{\text{measured}}}{T_{1}}\right) ^{2}+\left( \frac{\sigma _{\text{bottle}}}{\tau _{\text{bottle}}}\right) ^{2}+\left( \frac{\sigma _{\text{pulse}}}{T_{\text{pulse}}}\right) ^{2}+\left( \frac{\sigma_{\tau_3}}{\tau_{3}}\right) ^{2}+\left( \frac{\sigma_{T_{\rm 1,^3He}}}{T_{\rm 1,^3He}}\right) ^{2}
\end{align}

We see that the measurement of the neutron correlation function will require much more calibration and has more sources of error. This may require more measurements for improved statistics, but if all the other sources of error are well characterized a single measurement with a starting S/N of 100 would be required. 

\subsubsection{Transverse Relaxation (T$_{2}$) with RF gradient}

Another method to measure the correlation function would be to apply a $\frac{\pi }{2}$ pulse to the $^{3}$He and the neutrons and watch the FID decay from the scintillations of the spin dependent cross-section. Again, as with $^{3}$He this method is attractive because it eliminates the pulse loss systematic error, typically the largest source of error for $T_{1}~$measurements. For $T_{2}$ we have,
\begin{align}
\frac{1}{T_{2}}=\frac{1}{\tau _{\text{bottle}}}+\frac{1}{T_{\rm 2,rf\,gradient}}+\frac{1}{T_{\rm n,^3He}}+\frac{1}{T_{\rm 2,^3He}}
\end{align}

As the spins decay the oscillating signal from the spin-dependent cross section becomes a constant background signal. The total integrated background and signal can be used as a normalization. $T_{\rm 2,^3He},$ is the relaxation of $^3$He, this can be directly calculated from the correlation function measured during the $^{3}$He measurements. Errors would propagate like the previous cases. 

\subsubsection{Measuring the B$^{2}$ shift of neutrons}

\begin{figure}
\begin{center}
\includegraphics[width=.7\textwidth]{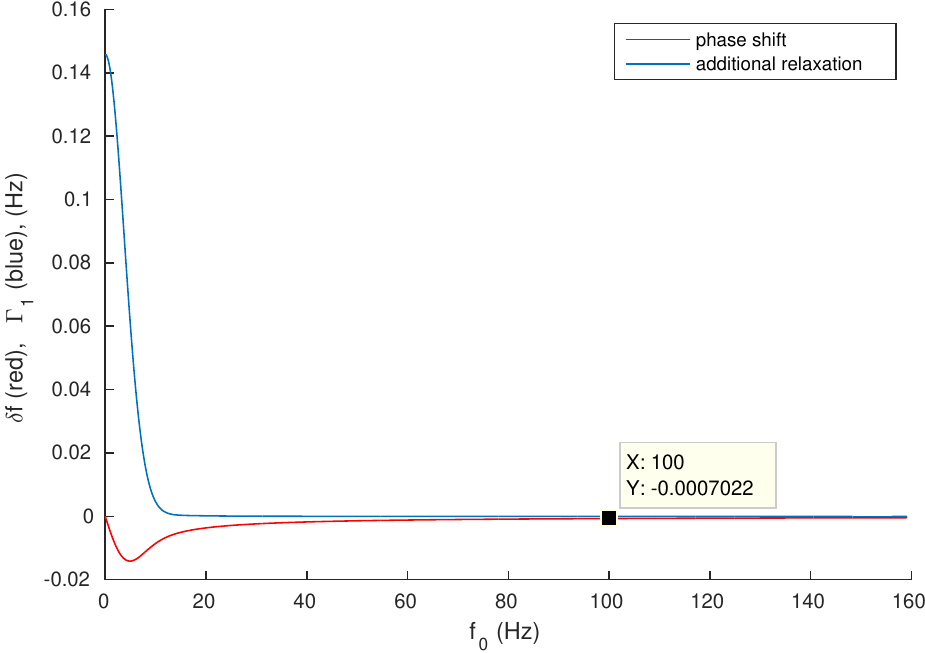}
\end{center}
\caption{Plot of the $B^2$ shift of neutrons in a $10\,{\rm \mu G\, cm^{-1}}$ gradient along the 40 cm cell direction}
\label{fig:neutronBsquared}
\end{figure}

Unlike the case for $^{3}$He, when we are measuring neutrons we are directly measuring the relative phase of the neutrons to the $^{3}$He. It should be possible to measure the frequency shift due to the additional $B$ field gradients. This possibility is shown in the Fig.~\ref{fig:neutronBsquared}, and where it is seen with $10~\mu G/cm$ gradient we find a shift of $7\times 10^{-4}~\mathrm{Hz}$, easily detectable in the SOS apparatus. 

Regardless of the ability to measure this shift I suspect it will be an important exercise, it is very attractive because it is a direct measurement of the imaginary part of the correlation function. There would be no model dependence. Furthermore this would be testing the measurement of a frequency shift, and could provide critical experience in measuring frequency shifts of UCN-$^{3}$He samples in an agile apparatus.

\subsection{Conclusion on correlation function measurements}

\begin{figure}
\begin{center}
\includegraphics[width=.7\textwidth]{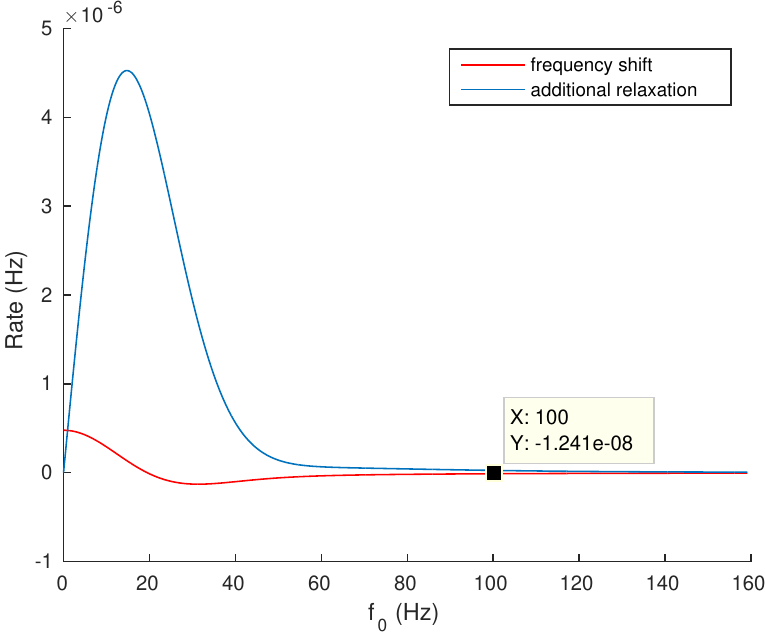}
\end{center}
\caption{Plot of the linear-in-E frequency shift of neutrons in a E-field of $75\,{\rm kV\,cm^{-1}}$ and a $10\,{\rm\mu G\,cm^{-1}}$ gradient field along the 40 cm cell direction}
\label{fig:neutronEB}
\end{figure}

In principle the SNS nEDM apparatus can make all the measurements that the SOS apparatus can make. For $^{3}$He it must use the same techniques to measure the correlation function.  However for UCNs there is a choice: either use correlation function or it can directly measure the false frequency shift (see Eq.~\ref{eq:dw}), by applying a $10~\mu G/cm$ gradient in presence of an $E$ field (such gradient will raise the false frequency shift while keeping $T_{2}$ long enough for an adequate measurement).  The numerical values of the measured effects for this two cases can be seen on  figure \ref{fig:neutronEB} (neutron false EDM) and figure \ref{fig:neutronBsquared} (neutron false EDM).

One can see that for the same gradient, the false edm frequency shift, $\delta\omega_{\mathrm{EB}}$, is $10^{-4}$ times smaller than the $\delta\omega_{B^2}$, making it more difficult to measure precisely. Furthermore it suffers form the same problem as the $B^{2}$ shift in that the coils must be centered to good precision. There will be additional systematic frequency shifts due an offset in the $B_{0}$ and gradient field coils, it will be distinguishable from $\delta\omega_{\mathrm{EB}}$ by reversing the $B~$gradient fields.

\section{Measuring neutron longitudinal relaxation time $T_{\rm 1,n}$ \label{appen:T1neutronMeasure}}

As mentioned in Sec.~\ref{sec:cellTesting}, the neutron longitudinal relaxation time $T_{\rm 1,n}$ can be measured by observing the scintillation light when leaving both the neutron and $^3$He spins aligned with $B_0$. Here we are discussing the case of no applied magnetic field gradients. In this case, the angle between the $^3$He and neutron polarizations is zero, and thus the scintillation light is given by Eq.~\ref{eq:FIDsignal} with $\phi(t) = \omega_{\rm free}t_i + \phi_0 = 0$ and replacing $T_{\rm 2,tot}$ with $T_{\rm 1,tot}$, where $ T_{\rm 1,tot}^{-1}= T_{\rm 1,n,walls}^{-1} + T_{\rm 1,^3He,walls}^{-1}$ (since the $T_{\rm 1,gradients}$ for the neutron and $^3$He are extremely long). That is, we expect the light signal to be:
\begin{eqnarray}
\frac{y(t_i)}{\Delta t} &=& I_0 {\rm e}^{-t_i/\tau_{\rm tot}} \left[1-F {\rm e}^{-t_i\left(T_{\rm 1,n,walls}^{-1} + T_{\rm 1,^3He,walls}^{-1}\right)} \right] + R_{\rm BG} \label{eq:T1neutronLight} \\
&=&  I_0 {\rm e}^{-t_i/\tau_{\rm tot}} - I_0 F {\rm e}^{-t_i\left(\tau_{\rm tot}^{-1}+T_{\rm 1,n,walls}^{-1} + T_{\rm 1,^3He,walls}^{-1}\right)}+ R_{\rm BG} \\
&=&  N_0 \frac{\epsilon_3}{\tau_3}\left[ \left(\frac{\epsilon_\beta \tau_3}{\tau_\beta\epsilon_3}+1 \right){\rm e}^{-t_i/\tau_{\rm tot}} - P_3 P_{\rm n} {\rm e}^{-t_i\left(\tau_{\rm tot}^{-1}+T_{\rm 1,n,walls}^{-1} + T_{\rm 1,^3He,walls}^{-1}\right)} \right]+ R_{\rm BG} \;.
\end{eqnarray}
This is a sum of two exponential decays with different time constants. As mentioned in Sec.~\ref{sec:cellTesting}, $\tau_{\rm walls}$ and $T_{\rm 1,^3He,walls}$ will be determined independently with different measurements. Fitting for the two time constants, and scanning the $x_3$ used, should allow $T_{\rm 1,^3He,walls}$ to be determined.

\end{appendices}

\bibliographystyle{apsrev4-1}
\footnotesize
\bibliography{SOSA_document 2019-09-09/SOSA_2017_document}

\begin{thebibliography}{9}%
\makeatletter
\providecommand \@ifxundefined [1]{%
 \@ifx{#1\undefined}
}%
\providecommand \@ifnum [1]{%
 \ifnum #1\expandafter \@firstoftwo
 \else \expandafter \@secondoftwo
 \fi
}%
\providecommand \@ifx [1]{%
 \ifx #1\expandafter \@firstoftwo
 \else \expandafter \@secondoftwo
 \fi
}%
\providecommand \natexlab [1]{#1}%
\providecommand \enquote  [1]{``#1''}%
\providecommand \bibnamefont  [1]{#1}%
\providecommand \bibfnamefont [1]{#1}%
\providecommand \citenamefont [1]{#1}%
\providecommand \href@noop [0]{\@secondoftwo}%
\providecommand \href [0]{\begingroup \@sanitize@url \@href}%
\providecommand \@href[1]{\@@startlink{#1}\@@href}%
\providecommand \@@href[1]{\endgroup#1\@@endlink}%
\providecommand \@sanitize@url [0]{\catcode `\\12\catcode `\$12\catcode `\&12\catcode `\#12\catcode `\^12\catcode `\_12\catcode `\%12\relax}%
\providecommand \@@startlink[1]{}%
\providecommand \@@endlink[0]{}%
\providecommand \url  [0]{\begingroup\@sanitize@url \@url }%
\providecommand \@url [1]{\endgroup\@href {#1}{\urlprefix }}%
\providecommand \urlprefix  [0]{URL }%
\providecommand \Eprint [0]{\href }%
\providecommand \doibase [0]{http://dx.doi.org/}%
\providecommand \selectlanguage [0]{\@gobble}%
\providecommand \bibinfo  [0]{\@secondoftwo}%
\providecommand \bibfield  [0]{\@secondoftwo}%
\providecommand \translation [1]{[#1]}%
\providecommand \BibitemOpen [0]{}%
\providecommand \bibitemStop [0]{}%
\providecommand \bibitemNoStop [0]{.\EOS\space}%
\providecommand \EOS [0]{\spacefactor3000\relax}%
\providecommand \BibitemShut  [1]{\csname bibitem#1\endcsname}%
\let\auto@bib@innerbib\@empty
\bibitem [{\citenamefont {Ahmed}\ \emph {et~al.}(2019)\citenamefont {Ahmed}, \citenamefont {Alarcon}, \citenamefont {Aleksandrova}, \citenamefont {Bae{\ss}ler}, \citenamefont {Barron-Palos}, \citenamefont {Bartoszek}, \citenamefont {Beck}, \citenamefont {Behzadipour}, \citenamefont {Berkutov}, \citenamefont {Bessuille} \emph {et~al.}}]{ahmed2019new}%
  \BibitemOpen
  \bibfield  {author} {\bibinfo {author} {\bibfnamefont {M.}~\bibnamefont {Ahmed}}, \bibinfo {author} {\bibfnamefont {R.}~\bibnamefont {Alarcon}}, \bibinfo {author} {\bibfnamefont {A.}~\bibnamefont {Aleksandrova}}, \bibinfo {author} {\bibfnamefont {S.}~\bibnamefont {Bae{\ss}ler}}, \bibinfo {author} {\bibfnamefont {L.}~\bibnamefont {Barron-Palos}}, \bibinfo {author} {\bibfnamefont {L.}~\bibnamefont {Bartoszek}}, \bibinfo {author} {\bibfnamefont {D.}~\bibnamefont {Beck}}, \bibinfo {author} {\bibfnamefont {M.}~\bibnamefont {Behzadipour}}, \bibinfo {author} {\bibfnamefont {I.}~\bibnamefont {Berkutov}}, \bibinfo {author} {\bibfnamefont {J.}~\bibnamefont {Bessuille}},  \emph {et~al.},\ }\href@noop {} {\bibfield  {journal} {\bibinfo  {journal} {Journal of Instrumentation}\ }\textbf {\bibinfo {volume} {14}},\ \bibinfo {pages} {P11017} (\bibinfo {year} {2019})}\BibitemShut {NoStop}%
\bibitem [{\citenamefont {Redfield}(1957)}]{redfield}%
  \BibitemOpen
  \bibfield  {author} {\bibinfo {author} {\bibfnamefont {A.}~\bibnamefont {Redfield}},\ }\href@noop {} {\bibfield  {journal} {\bibinfo  {journal} {IBM Journal}\ }\textbf {\bibinfo {volume} {January}},\ \bibinfo {pages} {19} (\bibinfo {year} {1957})}\BibitemShut {NoStop}%
\bibitem [{\citenamefont {Lamoreaux}\ and\ \citenamefont {Golub}(2005)}]{Lam05}%
  \BibitemOpen
  \bibfield  {author} {\bibinfo {author} {\bibfnamefont {S.~K.}\ \bibnamefont {Lamoreaux}}\ and\ \bibinfo {author} {\bibfnamefont {R.}~\bibnamefont {Golub}},\ }\href {\doibase 10.1103/PhysRevA.71.032104} {\bibfield  {journal} {\bibinfo  {journal} {Phys. Rev. A}\ }\textbf {\bibinfo {volume} {71}},\ \bibinfo {pages} {032104} (\bibinfo {year} {2005})}\BibitemShut {NoStop}%
\bibitem [{\citenamefont {{Guillaume Pignol, Stephanie Roccia}}(2012)}]{pignol}%
  \BibitemOpen
  \bibfield  {author} {\bibinfo {author} {\bibnamefont {{Guillaume Pignol, Stephanie Roccia}}},\ }\href@noop {} {\bibfield  {journal} {\bibinfo  {journal} {Physical Review A}\ }\textbf {\bibinfo {volume} {84}},\ \bibinfo {pages} {042105(5)} (\bibinfo {year} {2012})}\BibitemShut {NoStop}%
\bibitem [{\citenamefont {Swank}\ \emph {et~al.}(2016)\citenamefont {Swank}, \citenamefont {Petukhov},\ and\ \citenamefont {Golub}}]{Swank2016}%
  \BibitemOpen
  \bibfield  {author} {\bibinfo {author} {\bibfnamefont {C.~M.}\ \bibnamefont {Swank}}, \bibinfo {author} {\bibfnamefont {A.~K.}\ \bibnamefont {Petukhov}}, \ and\ \bibinfo {author} {\bibfnamefont {R.}~\bibnamefont {Golub}},\ }\href {\doibase 10.1103/PhysRevA.93.062703} {\bibfield  {journal} {\bibinfo  {journal} {Phys. Rev. A}\ }\textbf {\bibinfo {volume} {93}},\ \bibinfo {pages} {062703} (\bibinfo {year} {2016})}\BibitemShut {NoStop}%
\bibitem [{\citenamefont {{Riccardo Schmid}}(2013)}]{riccardoThesis}%
  \BibitemOpen
  \bibfield  {author} {\bibinfo {author} {\bibnamefont {{Riccardo Schmid}}},\ }\emph {\bibinfo {title} {{New search for the Neutron Electric Dipole Moment using Ultracold Neutrons at the Spallation Neutron Source}}},\ \href@noop {} {Ph.D. thesis},\ \bibinfo  {school} {California Institute of Technology} (\bibinfo {year} {2013})\BibitemShut {NoStop}%
\bibitem [{\citenamefont {{Christopher Swank}}(2012)}]{swankThesis}%
  \BibitemOpen
  \bibfield  {author} {\bibinfo {author} {\bibnamefont {{Christopher Swank}}},\ }\emph {\bibinfo {title} {{An Investigation in the Dynamics of Polarized Helium-3 in Superuid Helium-4 for the Spallation Neutron Source (SNS) neutron-electric-dipole-moment (nEDM) experiment.}}},\ \href@noop {} {Ph.D. thesis},\ \bibinfo  {school} {North Carolina State University} (\bibinfo {year} {2012})\BibitemShut {NoStop}%
\bibitem [{\citenamefont {McGregor}(1990)}]{mcgregor}%
  \BibitemOpen
  \bibfield  {author} {\bibinfo {author} {\bibfnamefont {D.~D.}\ \bibnamefont {McGregor}},\ }\href@noop {} {\bibfield  {journal} {\bibinfo  {journal} {Physical Review A}\ }\textbf {\bibinfo {volume} {45}},\ \bibinfo {pages} {2631} (\bibinfo {year} {1990})}\BibitemShut {NoStop}%
\bibitem [{\citenamefont {{G. Pignol, M. Guigue, A. Petukhov, and R. Golub}}(2015)}]{pignol2015}%
  \BibitemOpen
  \bibfield  {author} {\bibinfo {author} {\bibnamefont {{G. Pignol, M. Guigue, A. Petukhov, and R. Golub}}},\ }\href@noop {} {\bibfield  {journal} {\bibinfo  {journal} {Physical Review A}\ }\textbf {\bibinfo {volume} {92}},\ \bibinfo {pages} {053407} (\bibinfo {year} {2015})}\BibitemShut {NoStop}%
\end{thebibliography}%

\end{document}